\documentclass[aip,reprint,amsfonts,amsmath,amssymb]{revtex4-1}

\usepackage{graphicx}
\usepackage{bm}

\usepackage{color}
\usepackage{soul}

\usepackage{listings}

\begin{document}

\title{A semiclassical Thomas-Fermi model to tune the metallicity of electrodes in molecular simulations}
\author{Laura Scalfi}
\affiliation{Sorbonne Universit\'{e}, CNRS, Physico-chimie des \'Electrolytes et Nanosyst\`emes Interfaciaux, PHENIX, F-75005 Paris}
\affiliation{R\'eseau sur le Stockage Electrochimique de l'Energie (RS2E), FR CNRS 3459, 80039 Amiens Cedex, France}
\affiliation{These authors contributed equally to this work}
\author{Thomas Dufils}
\affiliation{Sorbonne Universit\'{e}, CNRS, Physico-chimie des \'Electrolytes et Nanosyst\`emes Interfaciaux, PHENIX, F-75005 Paris}
\affiliation{R\'eseau sur le Stockage Electrochimique de l'Energie (RS2E), FR CNRS 3459, 80039 Amiens Cedex, France}
\affiliation{These authors contributed equally to this work}
\author{Kyle Reeves}
\affiliation{Sorbonne Universit\'{e}, CNRS, Physico-chimie des \'Electrolytes et Nanosyst\`emes Interfaciaux, PHENIX, F-75005 Paris}
\affiliation{R\'eseau sur le Stockage Electrochimique de l'Energie (RS2E), FR CNRS 3459, 80039 Amiens Cedex, France}
\author{Benjamin Rotenberg}
\affiliation{Sorbonne Universit\'{e}, CNRS, Physico-chimie des \'Electrolytes et Nanosyst\`emes Interfaciaux, PHENIX, F-75005 Paris}
\affiliation{R\'eseau sur le Stockage Electrochimique de l'Energie (RS2E), FR CNRS 3459, 80039 Amiens Cedex, France}
\author{Mathieu Salanne}
\affiliation{Sorbonne Universit\'{e}, CNRS, Physico-chimie des \'Electrolytes et Nanosyst\`emes Interfaciaux, PHENIX, F-75005 Paris}
\affiliation{R\'eseau sur le Stockage Electrochimique de l'Energie (RS2E), FR CNRS 3459, 80039 Amiens Cedex, France}
\email{mathieu.salanne@sorbonne-universite.fr}


\begin{abstract}
	Spurred by the increasing needs in electrochemical energy storage devices, the electrode/electrolyte interface has received a lot of interest in recent years. Molecular dynamics simulations play a proeminent role in this field since they provide a microscopic picture of the mechanisms involved. The current state-of-the-art consists in treating the electrode as a perfect conductor, precluding the possibility to analyze the effect of its metallicity on the interfacial properties. Here we show that the Thomas-Fermi model provides a very convenient framework to account for the screening of the electric field at the interface and differenciating good metals such as gold from imperfect conductors such as graphite. All the interfacial properties are modified by screening within the metal: the capacitance decreases significantly and both the structure and dynamics of the adsorbed electrolyte are affected. The proposed model opens the door for quantitative predictions of the capacitive properties of  materials for energy storage.

\end{abstract}

\maketitle


\section{Introduction}

The development of constant applied potential methods for simulating
electrochemical systems~\cite{siepmann1995a} has allowed to solve many
outstanding problems in physical electrochemistry, ranging from the origin of
supercapacitance in nanoporous  electrodes made of carbon~\cite{merlet2012a} or
even of Metal Organic Frameworks~\cite{bi2020a} to the understanding of the
dynamic aspects of metal surface hydration~\cite{limmer2013b}. These methods are
based on the use of an extended Hamiltonian, in which the electrode charges are
additional degrees of freedom that obey a constant potential constraint at each
simulation step~\cite{limmer2013b,scalfi2020a}. They allowed to partly alleviate
the main conceptual difficulty to represent the electrode-electrolyte interface
at the molecular scale, which is the need to account for the electronic
structure on the electrode side, while the electrolyte is usually better
simulated using classical force fields because it requires a sampling of the
configurational space beyond the reach of today's capabilities with \textit{ab
initio} calculations (see Ref.~\citenum{scalfi2020arXiv} for a recent review 
of classical molecular simulations of electrode-electrolyte interfaces). 

Despite these successes, the possibility to simulate realistic systems remains limited by the crudeness of the ``electronic structure'' model, since the electrode is treated as a perfect metal. It is however well known that the electronic response of different electrodes (\textit{e.g.} graphite \textit{vs.} gold) to the adsorption of a charge should strongly differ. This was shown in numerous analytical~\cite{kornyshev1978a,kornyshev1982a} or density functional theory (DFT)-based studies~\cite{luque2012a,kornyshev2013a}, but also more recently in an experimental study where strong differences in the confinement-induced freezing of ionic liquids were shown depending on the nature of the electrode~\cite{comtet2017a}. In the latter study, this effect was interpreted using analytical developments accounting for the {\it metallicity} of the system, in the framework of the Thomas-Fermi (TF) model~\cite{kaiser2017a}.  

Here we build upon these developments to implement a computational Thomas-Fermi electrode. The TF model~\cite{thomas1927a,fermi1927a} is based on a local density approximation of the free electron gas, limited to its kinetic energy, and it accounts for the screening of the electrostatic potential over a characteristic screening length. 
We consider  model electrodes with the gold structure and tunable metallicity, separated by either vacuum or a simple NaCl aqueous electrolyte. We show that both the total accumulated charge and its distribution within the electrode are strongly affected. Accounting for screening in the electrodes radically changes their response to the adsorption of the electrolyte, which results in noticeable differences in the structure of the liquid when a voltage is applied. Screening inside the metal should therefore be accounted for when simulating electrochemical interfaces, in applications ranging from supercapacitors to Li-ion batteries.


\section{The Thomas-Fermi electrode model}

\begin{figure}[hbt!]
\centering
\includegraphics[width=0.9\columnwidth]{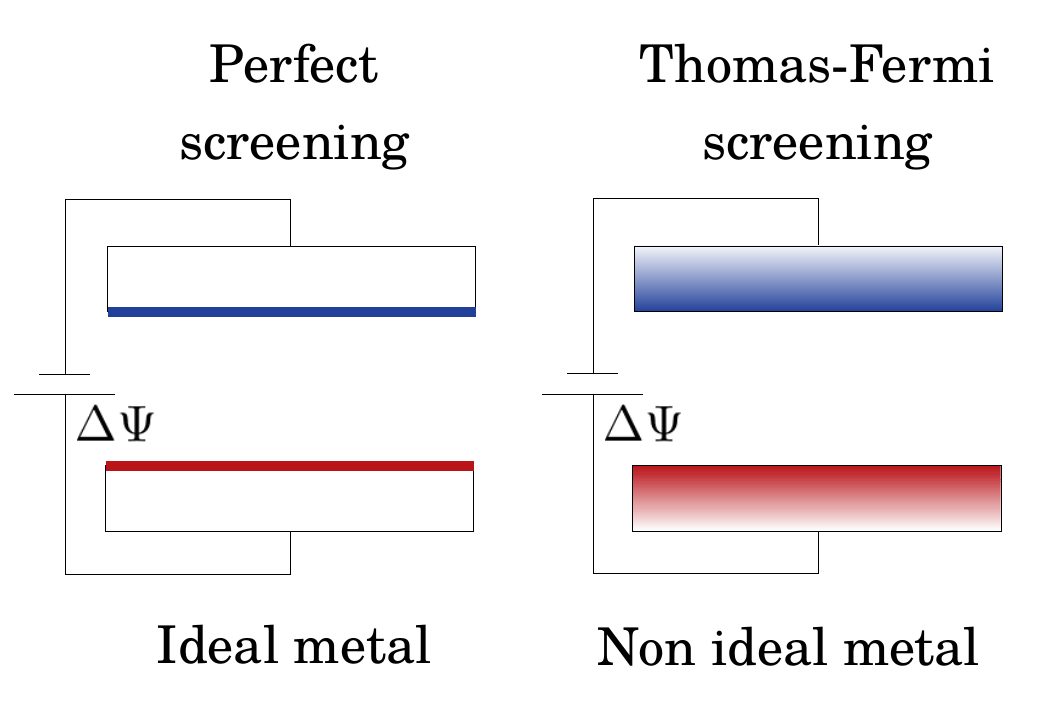}
\caption{Electrode polarization with different simulation methods. 
Constant potential simulations (left) correspond to a perfect screening of the charges, hence to the behavior of an ideal metal, whereas the Thomas-Fermi model introduces a screening length to account  for the imperfect screening of the charge in a non-ideal metal.}
\label{fig:scheme1}
\end{figure}

We consider an electrode composed of $N_{s}$ sites (here  these sites are positioned on the nuclei) with a number density $d$. Each atom $i$ has $Z$ valence electrons, and we introduce its partial charge $q_{i}$ as a dynamical variable accounting for the local excess of electrons. As shown schematically on Figure \ref{fig:scheme1}, in the currently available method the charges fluctuate in time to represent perfect metals. The partial charges are calculated at each simulation step in order to ensure that the potential is the same within the whole electrode~\cite{reed2007a}; when such an electrode is put in contact with an electrolyte the screening occurs within a thin layer at the surface only (note that supercapacitors are often simulated using constant charge setups, in which the vector $\{q_i\}_{i\in[1,N_{s}]}$ contains prescribed (usually identical) values for all the atoms of each electrode and does not vary with time, which does not correspond to a realistic electrode). Nevertheless, many electrode materials have a finite density of states available at the Fermi level. This was sometimes considered in the literature by computing a so-called quantum capacitance that accounts for the corresponding screening~\cite{kornyshev2013a,paek2015a}.

Here we propose to take these effects into account directly within classical molecular dynamics simulations, by employing the Thomas-Fermi model. It consists in a local density approximation of the energy of the valence electrons. The Thomas-Fermi functional for the kinetic energy reads
\begin{equation}
	U_{TF}[n(\mathbf{r})]=\int \dfrac{3}{10}\dfrac{\hbar^{2}}{m_e}(3\pi^{2})^{2/3}n(\mathbf{r})^{5/3} \mathrm{d}\mathbf{r}
	\; ,
\end{equation}
\noindent
where $n(\mathbf{r})$ is the local number density of electrons,
$m_e$ their mass and $\hbar$ Planck's constant. 
In order to obtain a practical description in molecular simulations,
we now express $n(\mathbf{r})$ as a sum over discrete atomic sites $i$, with
local densities $n_{i}=d\left[Z+\frac{q_{i}}{(-e)}\right]$, with $e$ the elementary charge. 
If the perturbation in the number of free charge carriers is small compared to the number of electrons, \textit{i.e.} $|q_i|\ll Ze$, we can expand the kinetic energy to second order in powers
of $q_i$ as
\begin{equation}\label{eq:utf}
U_{TF}= \dfrac{3}{5}N_{s}Z E_{F} + \dfrac{E_{F}}{(-e)} \sum_{i=1}^{N_s} q_{i}
+ \dfrac{l_{TF}^{2} d}{ 2\epsilon_{0}}\sum_{i=1}^{N_s} q_{i}^{2}
\end{equation}
\noindent where $E_{F}=\hbar^2k^2_F/2m_e$ is the Fermi level of a free-electron gas of density $Zd$ and $l_{TF}=\sqrt{\epsilon_{0}\hbar^{2}\pi^{2}/(m_ee^{2}k_{F})}$ is the Thomas-Fermi length of the material, with the corresponding Fermi wavevector defined by $k_{F}^{3}/3\pi^{2}=Zd$ and $\epsilon_{0}$ the vacuum permittivity.
The zeroth-order term is the total kinetic energy of an electron gas with $N_{s}Z$ electron (the total number of electrons in the system). The first order corresponds by definition to the chemical potential of the added/removed electrons (depending on the sign of $q_i$). The second order term, which is always positive and reaches its minimum when all the partial charges vanish corresponds to an energy penalty to induce non-homogeneous charge distributions.

Our system consists of two electrodes, hereafter named after their position in the simulation cell: left (L) and right (R). Their atom indices respectively range between $[1,N_L]$ and $[N_L+1,N_L+N_R]$, their Thomas-Fermi energies are noted $U_{TF}^{L}$ and $U_{TF}^{R}$, and they are held at potentials $\Psi_L$ and $\Psi_R=\Psi_L+\Delta\Psi$ where $\Delta\Psi$ is the applied voltage. We assume for simplicity that the electrodes are made of the same material, hence they have the same Fermi level at rest. The total energy of the system reads
\begin{equation}
E_{\rm tot}= K+U_C+U_{vdW}+U_{TF}^{L}+U_{TF}^{R}-\sum_{i=1}^{N_L}\Psi_Lq_i-\sum_{i=N_L+1}^{N_L+N_R}\Psi_Rq_i
\; ,
\label{eq:totalenergy1}
\end{equation}

\noindent where $K$ is the kinetic energy of the electrolyte, $U_{C}$ corresponds to the Coulombic interactions, $U_{vdW}$ describes the van der Waals interactions (given by a force field), while the last two terms account for the reversible work necessary to charge the electrode atoms. $U_C$ reads

\begin{equation}
U_{C}=\dfrac{1}{2}\iint \dfrac{\rho(\mathbf{r}) \rho(\mathbf{r'})}{4\pi\epsilon_{0}\vert \mathbf{r}-\mathbf{r}' \vert} \, {\rm d}\mathbf{r}\,{\rm d}\mathbf{r}'
\; , \label{eq:coulombenergy}
\end{equation}	
\noindent where the charge distribution $\rho(\mathbf{r})$ consists in a collection of $M$ point charges for the electrolyte and of $N=N_L+N_R$ atom-centered Gaussians (with width $\eta^{-1}$) representing the electrodes: 
\begin{equation}
		\rho(\mathbf{r})=\sum_{j=1}^M q_j\delta({\bf r}-{\bf r}_j)+\sum_{i=1}^N q_{i}\eta^{3}\pi^{-3/2} e^{-\eta^{2} \vert\mathbf{r}-\mathbf{r}_{i}\vert^{2} }
		\; ,
\end{equation}
with $\delta$ the Dirac function.
Note that in Eq.~\ref{eq:coulombenergy} the only self-energy to be included is the one due to the Gaussian charges. By injecting Eq.~\ref{eq:utf} into Eq.~\ref{eq:totalenergy1} and introducing $\Delta \Psi$, the total energy can be rewritten as
\begin{eqnarray}
\label{eq:Utot}
	E_{\rm tot} &=& K +  U_C + U_{vdW} +\dfrac{3}{5}NZE_{F}
	+\dfrac{l_{TF}^{2} d}{2\epsilon_{0}}\sum_{i=1}^N q_{i}^{2} \nonumber\\
	             &&-(\Psi_L+\frac{E_F}{e})\sum_{i=1}^Nq_i-\Delta\Psi\sum_{i=N_L+1}^N q_i
	             \; .
\end{eqnarray}

\noindent 
As detailed in Ref.~\citenum{scalfi2020arXiv}, in the absence of electrochemical reactions, we impose the electroneutrality constraint $\sum_{i=1}^{N}q_i=0$, so that the electrodes bear opposite charges and the corresponding term in the reversible work reduces to the usual $Q_{tot}\Delta\Psi$, with $Q_{tot}$ the total charge of the positive electrode.
As in the constant potential method neglecting the quantum nature of the electrons (corresponding to $l_{TF}=0.0$~\AA), the charges are treated as dynamic variables which are obtained at each time step of the simulation by enforcing the constant potential constraint $\partial E_{\rm tot} / \partial q_i = 0$ \citep{scalfi2020a,reed2007a}. Compared to this perfect metal case, the modifications of the algorithm are minimal and virtually don't add any computational cost.

Our approach, which involves fluctuating charges, may be related to the charge
equilibration model~\cite{nalewajski1984a,mortier1986a,rappe1991a}, in
particular to its extension to electrochemical systems proposed by Onofrio {\it
et al.}~\cite{onofrio2015a}. This method is based on two main chemical
quantities, the electronegativity $\chi$ and the hardness $H$ of each atomic
species. The self-consistent equations to solve are equivalent if we take $\chi
\sim E_F$ and $H \sim e^2 l_{TF}^2d/\epsilon_0$. However, these concepts, which
are related to those of electronic affinity and ionization
energy~\cite{buraschi2020a}, are rooted in the description of the electronic
properties of atoms and molecules, rather than that of bulk materials, which are
more naturally described in terms of band structure.  
The issue of starting from the correct reference state 
for (electro-)chemical potential equalization methods was 
already pointed out in Ref.~\citenum{york1996a}, where York and Yang
derived a fluctuating charge model from DFT for molecules
and underlined the difference between atomic and molecular reference states
to determine the electronegativities and hardnesses. More recently, a
 detailed discussion on the correspondance between constant potential electrode
models and the charge equilibration approach was provided in Ref.~\citenum{nakano2019a}.
Another physical model of electrodes was proposed~\cite{pastewka2011a}, in which the Hamiltonian is constructed in the tight-binding approximation.


\begin{figure*}[hbt!]
\centering
\includegraphics[width=0.8\textwidth]{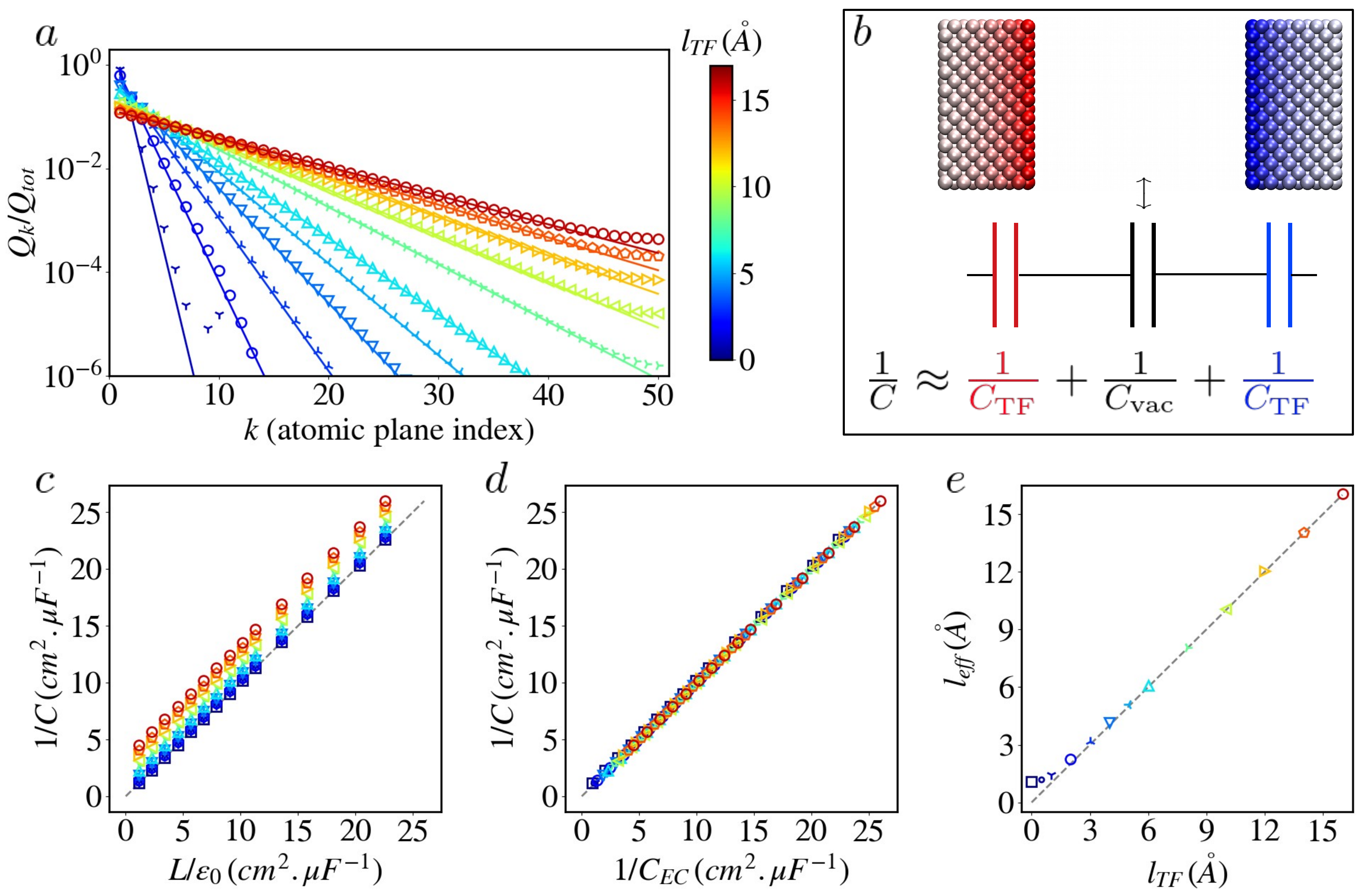}
	\caption{
		Empty Thomas-Fermi capacitor. All results correspond to a (100) gold-like electrode structure with 
    	$n=50$ atomic planes and $L=300$~\AA\ between the electrodes where not stated otherwise. 
    	Charges were computed by applying a voltage $\Delta\Psi=1$~V between the electrodes for different Thomas-Fermi lengths $l_{TF}$ ranging from 0.0 to 16.0~\AA, that are represented both by different symbols and by different colors indicated in the colorbar. 
    	(a) Total charge per plane on the positive electrode as a function of the position from the surface ($k$ is the index of the atomic plane), normalized by the total electrode charge $Q_{tot}$.
    	The symbols are simulated values for different Thomas-Fermi lengths $l_{TF}$ and the lines are the prediction of Eq.~\ref{eq:charge}. 
    	(b) Snapshot of the simulated system and its equivalent circuit representation corresponding to the capacitance obtained with the continuum theory (see text).
    	(c) Computed reciprocal capacitance as a function of the analytical predictions for perfect metals using $L_{vac}=L$, and (d) for Thomas-Fermi metals using Eq.~\ref{eq:prediction} with $L_{vac}=L-a$, for varying electrode spacing $L$ (between 10 and 200~\AA). (e) Effective length $l_{eff}$, defined in Eq.~\ref{eq:leff}, as a function of $l_{TF}$.
    	}
\label{fig:empty}
\end{figure*}

\section{Empty capacitor} \label{sec:validation}

As a first validation of our implementation, we study a model system composed of two planar (100) gold electrodes separated by a distance $L$ and held at a constant potential difference $\Delta\Psi=1$~V. Each electrode consists of $n$ atomic planes with an inter-spacing $a$ in the $z$ direction. We compare the simulated results against analytical predictions of the corresponding continuum model where the Poisson equation for the one dimensional potential $\Psi(z)$ reads $\Psi''(z)=l_{TF}^{-2}\Psi(z)$ inside each electrode and $\Psi''(z)=0$ between them. The total capacitance of the system is given by $C = Q_{tot} / \Delta \Psi$.

Assuming that the width of the material is large compared to the Thomas-Fermi length, the in-plane charge $Q_k$ at $z = k a$ ($k\in[1,n]$) can be expressed as
\begin{equation} \label{eq:charge}
\frac{Q_k}{Q_{tot}}=e^{-(k-1)a/l_{TF}} \left[1-e^{-a/l_{TF}} \right],
\end{equation}
Figure \ref{fig:empty}a shows a very good agreement between Eq.~\ref{eq:charge} and the simulation for large $l_{TF}$ values. Small deviations for large $z$ are due to the finite number of planes. 
The above exponentially decaying charge distribution inside the metal, due to the screening over the Thomas-Fermi length $l_{TF}$, results according to the continuous model in a capacitance per unit area
\begin{equation} \label{eq:prediction}
	\frac{1}{C_{EC}} = \frac{1}{C_{vac}} + \frac{2}{C_{TF}} = \frac{L_{vac}}{\epsilon_0 } + \frac{2l_{TF}}{\epsilon_0 }
\; ,
\end{equation}
with $C_{vac}=\epsilon_0/L_{vac}$ the theoretical capacitance per unit area for
perfect metallic electrodes ($l_{TF}=0$~\AA) separated by a vacuum slab of width
$L_{vac}$ and $C_{TF}=\epsilon_0/l_{TF}$ that for a single Thomas-Fermi
electrode. This result can be simply understood in terms of the equivalent
circuit (hence the subscript $C_{EC}$) illustrated in Figure~\ref{fig:empty}b,
with three capacitors in series (see Supplementary Section S1 for a discussion of the continuum descriptions and equivalent circuit models). 
As shown in Figure~\ref{fig:empty}c, the simulation results are consistent with the prediction of a linear relation between $1/C$ and $L/\epsilon_{0}$, where $L$ is the distance between the first atomic planes on each electrode, with a constant shift which increases with $l_{TF}$. 

However, the width of the vacuum slab between the electrodes is not exactly the
distance between the first atomic planes. Indeed, each atomic site is surrounded
by electrons, and the boundary between the free electron gas inside the
electrode and the vacuum~\cite{lang1973a} (the so-called ``Jellium
edge''~\cite{smith1989a}) is rather shifted half of the inter-plane distance
away from the electrode. Since this feature is present on both electrodes, the
actual vacuum slab width is more consistent with $L_{vac}=L-a$.
Figure~\ref{fig:empty}d shows that using this prescription,
Eq.~\ref{eq:prediction} provides a very good description of the simulated
capacitance $C$ over a wide range of distances between the electrodes and
Thomas-Fermi lengths, which confirms the consistency of the present classical
model to represent the charge distribution within the metal. The decay length of the charge inside the electrode coincide with $l_{TF}$ within 1~\% for all values $l_{TF}\gtrsim a$. The slight deviations from the predictions of the continuous theory can be analyzed by introducing an effective length $l_{eff}$ from the measured capacitance as
\begin{equation}\label{eq:leff}
	\frac{1}{C} = \frac{L - a}{\epsilon_0} + \frac{2l_{eff}}{\epsilon_0}
	\; .
\end{equation}
The results obtained for various $l_{TF}$ at fixed $L$, illustrated in
Figure~\ref{fig:empty}e, indicate that this effective length deviates from the
Thomas-Fermi length only when the latter becomes comparable to the atomic
details of the electrodes (interplane and interatomic distances, width of the
Gaussian distributions). An additional test was performed by adding a single
charge at various distances between the electrodes and comparing the energy of
the system to an approximate analytical expression~\cite{kaiser2017a}. The
results, which are provided in Supplementary Section S2, also show a good agreement over a broad range of $l_{TF}$ values.

\section{Impact of the Thomas-Fermi length on the electrochemical interface properties}

\begin{figure*}[hbt!]
\centering
\includegraphics[width=\textwidth]{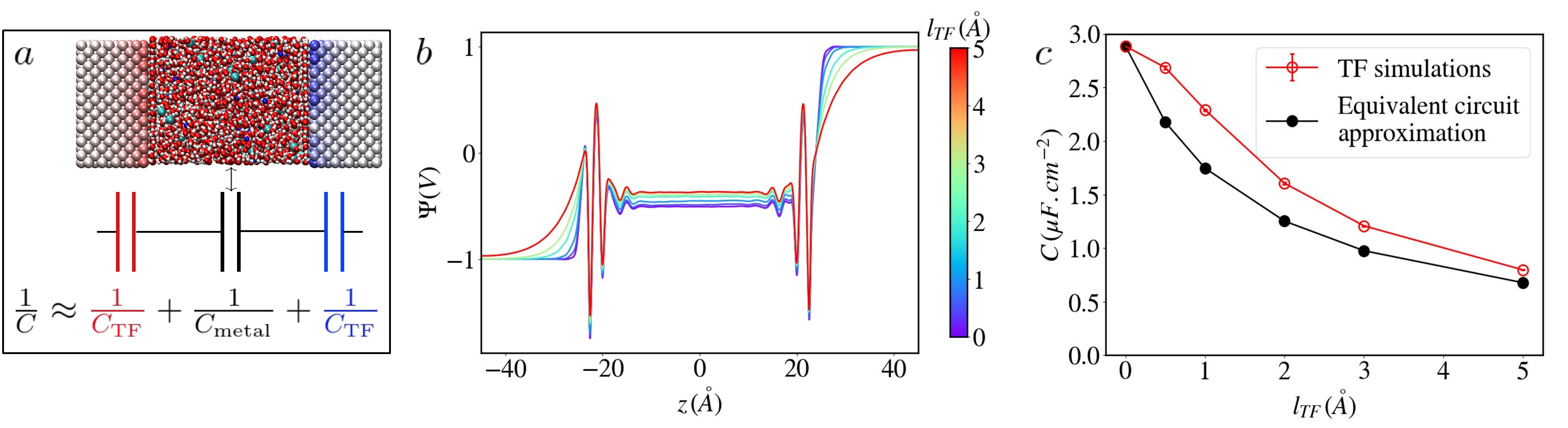} 
	\caption{The capacitance decreases significantly with Thomas-Fermi
length. (a) Snapshot of the simulated system and its equivalent circuit
representation, where $C_{\rm metal}$ stands for the capacitance
computed for the perfect metal simulation. (b) Poisson potential across the
simulation cell for a system made of two (100) gold-like electrodes in contact
with a NaCl aqueous solution. The applied voltage is 2~V and different $l_{TF}$
values ranging from 0.0 to 5.0~\AA\ are represented by different colors
indicated in the colorbar. The screening of the potential inside the electrodes
increases markedly with $l_{TF}$. (c) Variation of the capacitance with $l_{TF}$. The results from the simulations are compared with the equivalent circuit approximation. Error bars are extracted from the standard error of the charge distribution corrected for sample correlations.}
\label{fig:goldcapacitance}
\end{figure*}

In order to understand the impact of screening inside the metal on the
properties of electrode/electrolyte interfaces, we study a system consisting of
two (100) gold-like electrodes in contact with an aqueous solution of NaCl (with
concentration 1~mol~L$^{-1}$), illustrated in Figure~\ref{fig:goldcapacitance}a. 
The TF length $l_{TF}$  
was systematically varied from 0.0 to 5.0~\AA\ in order to switch from a perfect metal to typical semi-metallic conditions (estimations yield typical values of 0.5~\AA\ for platinum, 1.5~\AA\ for doped silicon and 3.4~\AA\ for graphite~\cite{comtet2017a}). Simulations were performed for voltages $\Delta\Psi=$0, 1 and 2~V between the two electrodes.

As a first illustration of the impact of screening on the electrochemical interface, we compute the Poisson potential across the cell. The results for an applied potential of 2~V are displayed on Figure~\ref{fig:goldcapacitance}b. We observe a very different pattern inside the electrode depending on $l_{TF}$: for the perfect metal the applied potential is reached at positions corresponding to the first atomic plane, while for the TF model we clearly see the desired effect of field penetration with an exponential decay inside the electrode.

Figure~\ref{fig:goldcapacitance}c shows that the integral capacitance decreases
significantly with $l_{TF}$ (note that it remains constant between 1 and 2~V,
see Supplementary Figure S2). The effect is already non negligible for $l_{TF}=0.5$~\AA~(which is representative of many real metals) since the capacitance is 7~\% smaller than the one of the perfect metal; it is even more pronounced in the semi-metallic r\'egime. This can be understood by noting that the TF length varies as the inverse square-root of the number of available states at the Fermi level. In a perfect metal, the number of accessible states is infinite, so that the only resistance to charging arises from the Coulombic energy. In contrast, the TF model results in an additional energy penalty for increasing the surface charge, described by the quadratic term in Eq.~\ref{eq:Utot}. 

As for the empty capacitor, it is possible to estimate the capacitance from the
value for the perfect metal $C_{\rm metal}$ using the equivalent circuit
depicted on Figure~\ref{fig:goldcapacitance}a (see Supplementary Section
S1). This approach, used for example by Gerischer to interpret experimental data~\cite{gerischer1985a}, has been applied in many simulation works where the additional term due to the screening was computed using DFT and therefore called ``quantum capacitance'', while the perfect metal capacitance was computed using either a mean-field theory~\cite{kornyshev2013a} or  molecular dynamics~\cite{pak2013a}.  Nevertheless, it neglects the interplay between the electronic structure of the electrode and the ionic structure of the adsorbed electrolyte. This coupling is self-consistently taken into account in our model, which therefore provides a perfect framework to test this approximation. As can be seen on Figure \ref{fig:goldcapacitance}c, the equivalent circuit approximation underestimates rather significantly the real capacitance (by 20 to 30~\%). 

\begin{figure*}[hbt!]
\centering
\includegraphics[width=0.8\textwidth]{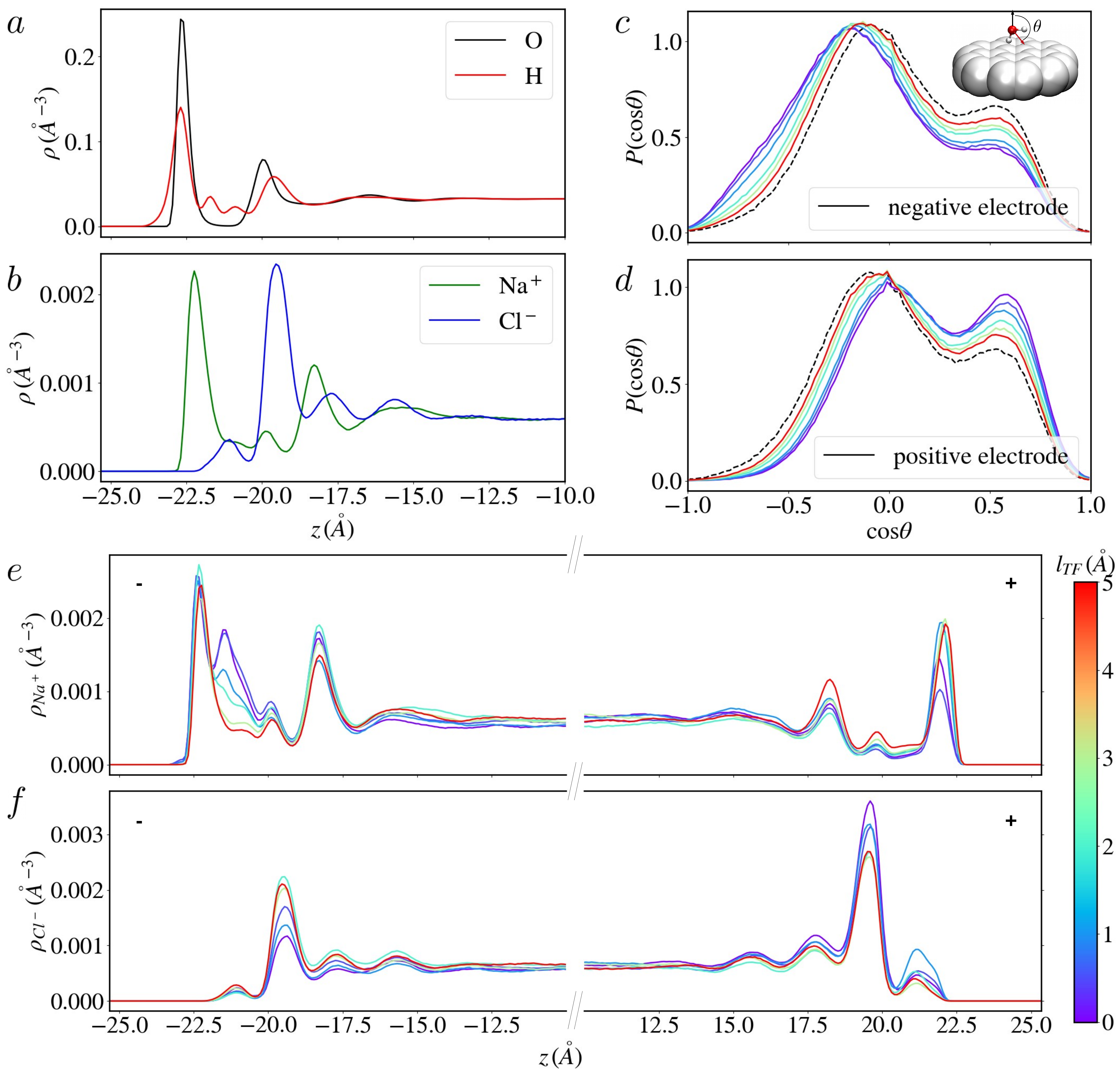}
	\caption{The structure of the electrochemical interface depends on the
Thomas-Fermi length at finite voltages. (a,b) Atomic density profiles for the O,
H, Na$^+$ and Cl$^-$ atoms near the electrode at null potential for
$l_{TF}=0.0$~\AA\ (the profiles are the same for the other $l_{TF}$ values as
shown on Supplementary Figure S3). Note that in the case of H atoms the profile is divided by two to facilitate the comparison with O atoms. (c,d) Distribution of the adsorbed water molecules orientation with respect to the vector normal to the electrode surfaces for an applied potential of 2~V for the whole range of simulated $l_{TF}$ indicated by the colorbar; the distribution for 0~V and $l_{TF}=0.0$~\AA\ is also reported (black dashed lines) as a reference. (e,f) Atomic density profiles for the Na$^+$ and Cl$^-$ ions for an applied potential of 2~V for the whole range of simulated $l_{TF}$ indicated by the colorbar. The negative (positive) electrode is located at negative (positive) $z$.}
\label{fig:structure}
\end{figure*}

At null voltage, the average structure of the liquid does not vary significantly
with $l_{TF}$ (see Supplementary Figure S3). As shown on Figure
\ref{fig:structure}a-b, it is characterized by several adsorption layers, mainly
consisting of water molecules. By computing the distribution of the angle
$\theta$ between the vector normal to the surface and the water dipole (see
dashed black curve on Figure \ref{fig:structure}c-d) or the O-H bonds (see
Supplementary Figure S4) for molecules in the first adsorbed layer, we observe that they mostly lie in a plane parallel to the surface or with one H atom pointing away from the surface. A small population is oriented towards the surface, which results in a small shoulder on the H atoms atomic density profiles. 

The ions have different adsorption profiles: the Na$^+$ density is characterized
by a large peak located close to the one of O atoms, so that they can be
considered to belong to the first layer, while the Cl$^-$ ions are located
further away from the electrode surface. Their profile displays a small pre-peak
in the region where the water density is very low and a peak with a larger
intensity located in the second hydration layer. Once a potential is applied,
the liquid mainly responds on the two electrodes through (i) a stronger
orientation of the water molecules towards/away from the negative/positive
electrode as shown on Figure \ref{fig:structure}c-d and Supplementary
Figures S4 and S5, and (ii) the appearance of a new adsorption peak for the Na$^+$ ions near the negative electrode (Figure \ref{fig:structure}e) and an increase of the pre-peak intensity in the Cl$^-$ density profiles on the positive electrode side (Figure \ref{fig:structure}f). In all cases, the modifications in the structure depend strongly on $l_{TF}$. This shows that depending on the type of material, we can expect all the electrochemical double-layer properties to change markedly with the nature of the chosen electrode.

\begin{figure}[hbt!]
\centering
\includegraphics[width=\columnwidth]{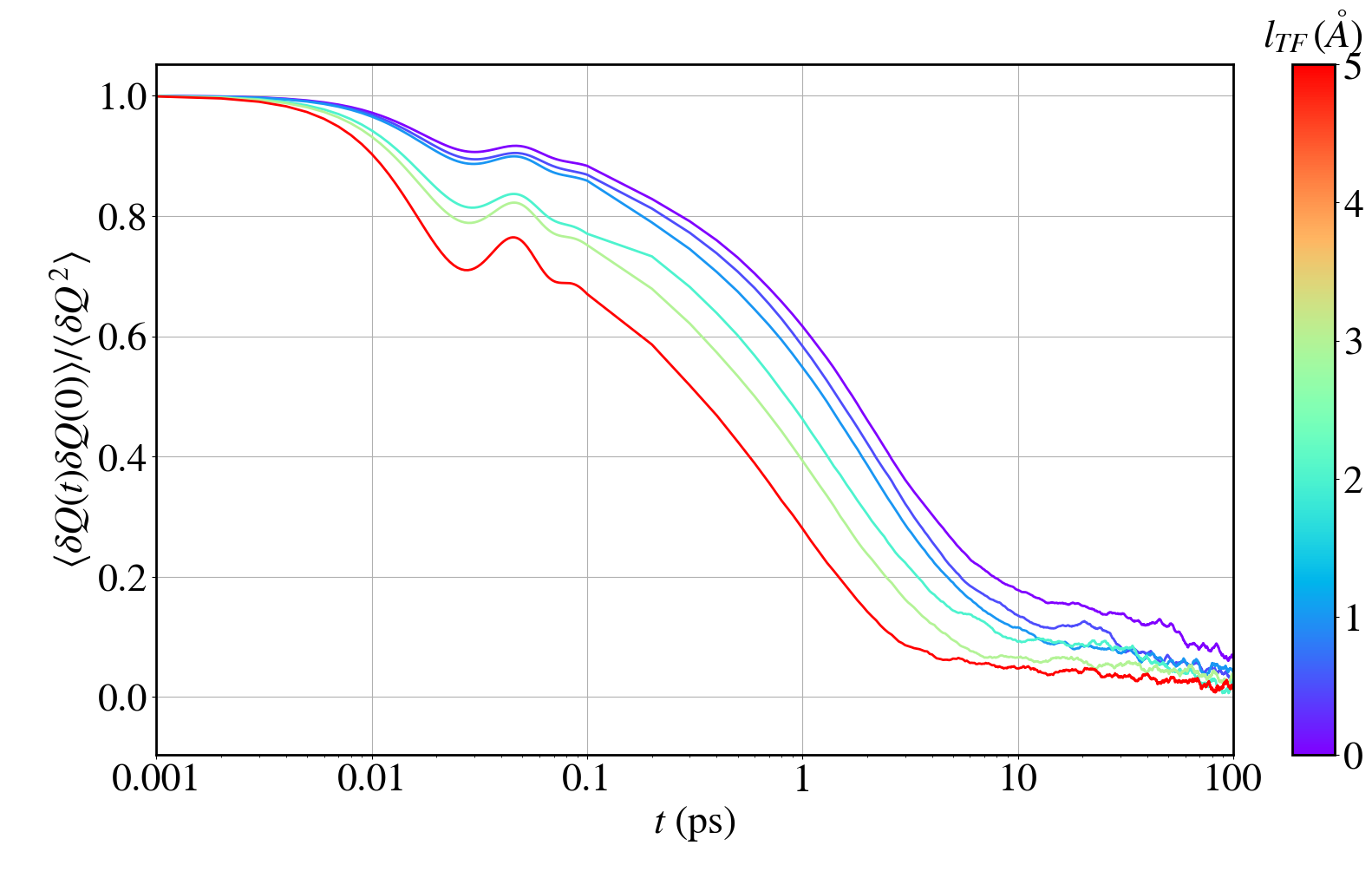}
	\caption{
		The relaxation of the electrode charge indicates a faster dynamics of the interfacial electrolyte near screened metals. Normalized auto-correlation function of the total charge at null potential for varying $l_{TF}$ values ranging from 0.0 to 5.0~\AA \, indicated by the colorbar.}
\label{fig:autocorrelation}
\end{figure}

Dynamical properties are particularly important for electrochemical applications. They control for example the power delivered by an energy storage device. The equilibrium fluctuations of the electrode charge at 0~V, which reflect the linear response to a small applied voltage, are shown on Figure \ref{fig:autocorrelation} for the various $l_{TF}$. An increased screening yields faster dynamics for the relaxation of the electrochemical double-layer. Such a difference was somewhat unexpected given that the systems at null potential have on average the same structural features, but it can be qualitatively understood as the result of weaker interactions with the more diffuse charges induced within the electrode. This means that the dynamics do not only depend on the nature of the electrolyte, but also on the electronic structure of the electrode material.  


\section{Conclusion}
Understanding the electrode/electrolyte interface is a prerequisite not only for the design of more efficient energy storage devices~\cite{salanne2016a}, but also for understanding wetting phenomena involved in lubrication or heterogeneous catalysis~\cite{carrasco2012a}. Although in the past decades molecular simulations have provided many insights on the structure of the electrochemical double-layer, they still fail at predicting quantitatively many experimental quantities, such as the variation of the differential capacitance with the applied voltage~\cite{fedorov2014a}. This is particularly true in the case of carbon materials, due to their complex electronic structure properties that deviate largely from the ones of typical metals. Many intriguing experimental observations, such as the capillary freezing of ionic liquids  confined between metallic surfaces~\cite{comtet2017a} or the emergence of longer-than-expected electrostatic screening lengths in concentrated electrolytes~\cite{gebbie2013a,smith2016a}, remain to be explained quantitatively. The Thomas-Fermi model, by allowing to tune the metallicity of the electrode using a single parameter (and without introducing additional computational costs) should lead to a more accurate understanding of the interfacial properties of such electrodes using molecular simulations.
The extension of this work to complex  materials such as nanoporous carbons will require additional efforts, in order to take into account the effect of the local environment of each atom on its electronic response. In that case, it might be relevant to sacrifice some of the simplicity of the TF model by including atom-specific or even bond-specific terms in the energy, following \emph{e.g.} the split charge equilibration approach~\cite{nistor2006a,nistor2009a}. In this context, the present work suggests that it could be possible to determine the associated parameters from a simplified representation of the underlying electronic density. 

\section*{Appendix : Simulation details}
The TF electrode model was implemented in the molecular dynamics code
MetalWalls~\cite{marinlafleche2020a}. All simulations were run using a matrix
inversion method\cite{scalfi2020a} to enforce both the constant potential and
the electroneutrality constraints on the charges. Electrode atoms have a
Gaussian charge distribution of width $\eta^{-1} = 0.56$~\AA\ centered on zero
and the Thomas-Fermi length $l_{TF}$ ranges from 0.0 to 16.0~\AA\ for the empty capacitor and from 0.0 to 5.0~\AA\ in the presence of aqueous NaCl electrolyte. Two-dimensional boundary conditions were used with no periodicity in the $z$ direction using an accurate 2D Ewald summation method to compute electrostatic interactions. A cutoff of 17.0~\AA\ was used for both the short range part of the Coulomb interactions and the intermolecular interactions. For the latter we used the truncated shifted Lennard-Jones potential. The box length in both the $x$ and $y$ directions was $L_{x} = L_y = 36.630$~\AA\ with 162 atoms per atomic plane. The structure is face-centered cubic with a lattice parameter of $4.07$~\AA\ and a separation between planes $a = 2.035$~\AA\ in the (100) direction (the atomic density $d$ is 0.59~$\cdot$~10$^{29}$~m$^{-3}$). The empty capacitors have 50 planes per electrode whereas the electrochemical cells have 10 (leading to a total of 1620 atoms per electrode). In the latter case, the electrolyte is composed of 2160 water molecules, modeled using the SPC/E force field~\cite{berendsen1987a}, and 39 NaCl ion pairs. The Lennard-Jones parameters for Na$^+$ and Cl$^-$ were taken from Ref.~\citenum{dang1995c} and the ones for the electrode atoms from Ref.~\citenum{berg2017a}; the Lorentz-Berthelot mixing rules were used.
The simulation boxes were equilibrated at constant atmospheric pressure for 500~ps by applying a constant pressure force to the electrodes with $l_{TF}=0.0$~\AA\ then the electrodes separation was fixed to the equilibrium value (for which the density in the middle of the liquid slab is equal to its bulk value) $L=$~56.8~\AA. The simulations were run at 298~K with a timestep of 1~fs. Each system was run for at least 8~ns.


\section*{Acknowledgements}
\noindent The authors thank M. Sprik, P. A. Madden, L. Bocquet and B. Coasne for useful discussions. This project has received funding from the European Research Council  under the European Union's Horizon 2020 research and innovation programme (grant agreement No. 771294). This work was supported by the French National Research Agency (Labex STORE-EX, Grant  ANR-10-LABX-0076, and project NEPTUNE, Grant  ANR-17-CE09-0046-02).


\section*{Supplementary information}

\setcounter{section}{0}
\setcounter{figure}{0}
\setcounter{equation}{0}

\renewcommand\theequation{S\arabic{equation}}
\renewcommand\thefigure{S\arabic{figure}}
\renewcommand\thesection{S\arabic{section}}


\section{Continuum description and equivalent circuit models}

\noindent
As discussed in the main text, the capacitance of the electrochemical cell comprising the electrodes, the electrolyte and the interfaces between them is often analysed in term of a simple model based on capacitors in series. Considering the symmetry of the problem, the mean-field Poisson potential $\Psi$ only depends on the position $z$ in the direction perpendicular to the electrodes and satisfies one of the following equations:

\begin{itemize}
\item Thomas-Fermi: $\Psi''(z) = \dfrac{1}{l_{TF}^2} \Psi(z)$ inside the electrodes described by the Thomas-Fermi model, with $l_{TF}$ the Thomas-Fermi length of the material.
\item Debye-H\"uckel: $\Psi''(z) = \dfrac{1}{\lambda_D^2} \Psi(z)$, with $\lambda_D=\left( \frac{e^2}{\epsilon_0\epsilon_r k_BT}\sum_i c_i z_i^2\right)^{-1/2}$ the Debye screening length for a dilute electrolyte, with $e$ the elementary charge, $k_BT$ the thermal energy, $\epsilon_r$ the relative permittivity of the solvent, $c_i$ the concentration of ionic species $i$ and $z_i$ its valency. This equation is the linearized version of the more general Poisson-Boltzmann equation and is only valid for small applied voltages.
\item Poisson: $\Psi''(z) = -\frac{\rho_{q}(z)}{\epsilon_0}$ in the more general case, with $\rho_{q}$ the charge density which can be obtained from the density profiles of all species (including the contributions of the O and H of water in the aqueous systems considered in the present work).
\end{itemize}

These equations need to be solved in the various regions of the system, with appropriate continuity equations at the boundaries between them and overall boundary conditions $\lim_{z\to-\infty}\Psi(z)=\Psi_L$ and $\lim_{z\to+\infty}\Psi(z)=\Psi_R$, with $\Psi_L$ and $\Psi_R$ the potentials of the left and right electrodes, respectively. The capacitance of the system, defined as the ratio between the charge accumulated inside the electrodes and the voltage $\Delta\Psi=\Psi_R-\Psi_L$, can then be expressed as
\begin{equation}
\label{eq:capaseries}
\dfrac{1}{C} = \sum_k \dfrac{1}{C_k}
\end{equation}
with $C_k$ the capacitance corresponding to each region $k$. This rule allows to determine the relative contribution of each region as a function of the composition (electrode material, electrolyte) and geometry (distance between the electrodes) of the system.

In both the Thomas-Fermi electrode and the Debye-H\"uckel electrolyte, the potential behaves as a sum of two exponentials. If the width of the corresponding regions (electrode slab or electric double layer) are large compared to the associated screening lengths, then only one exponential contributes and the capacitance is given by $C_{TF}=\dfrac{\epsilon_0}{l_{TF}}$ (analogous to the ``quantum capacitance'' sometimes introduced to capture the contribution of the electrode) and $C_{DH}=\dfrac{\epsilon_0}{\lambda_{D}}$, respectively. For an empty capacitor, the vacuum slab between the electrodes corresponds to a capacitance $C_{vac}=\dfrac{\epsilon_0}{L_{vac}}$, with $L_{vac}$ the width of the slab, \emph{i.e.} the distance between the surface of the electrodes (which slightly differs from the difference between the position of the first atomic planes on each surface, as explained in the main text).

For the pure water case, one should distinguish the contribution of the bulk dielectric liquid, with capacitance $C_{bulk}=\dfrac{\epsilon_0\epsilon_r}{L_{bulk}}$, and that of the structured layers at the interface, which can be associated with an effective interfacial capacitance $C_{w,int}$. This interfacial contribution of water is also present in the aqueous electrolyte, and the ionic contribution may be estimated with $C_{DH}$ in the dilute and low potential limits, from the non-linear Poisson-Boltzmann capacitance, or directly from the ionic concentration profiles from molecular simulations.

\section{Single charge between two electrodes}

\begin{figure}[hbt!]
\centering
\includegraphics[width=1.0\columnwidth]{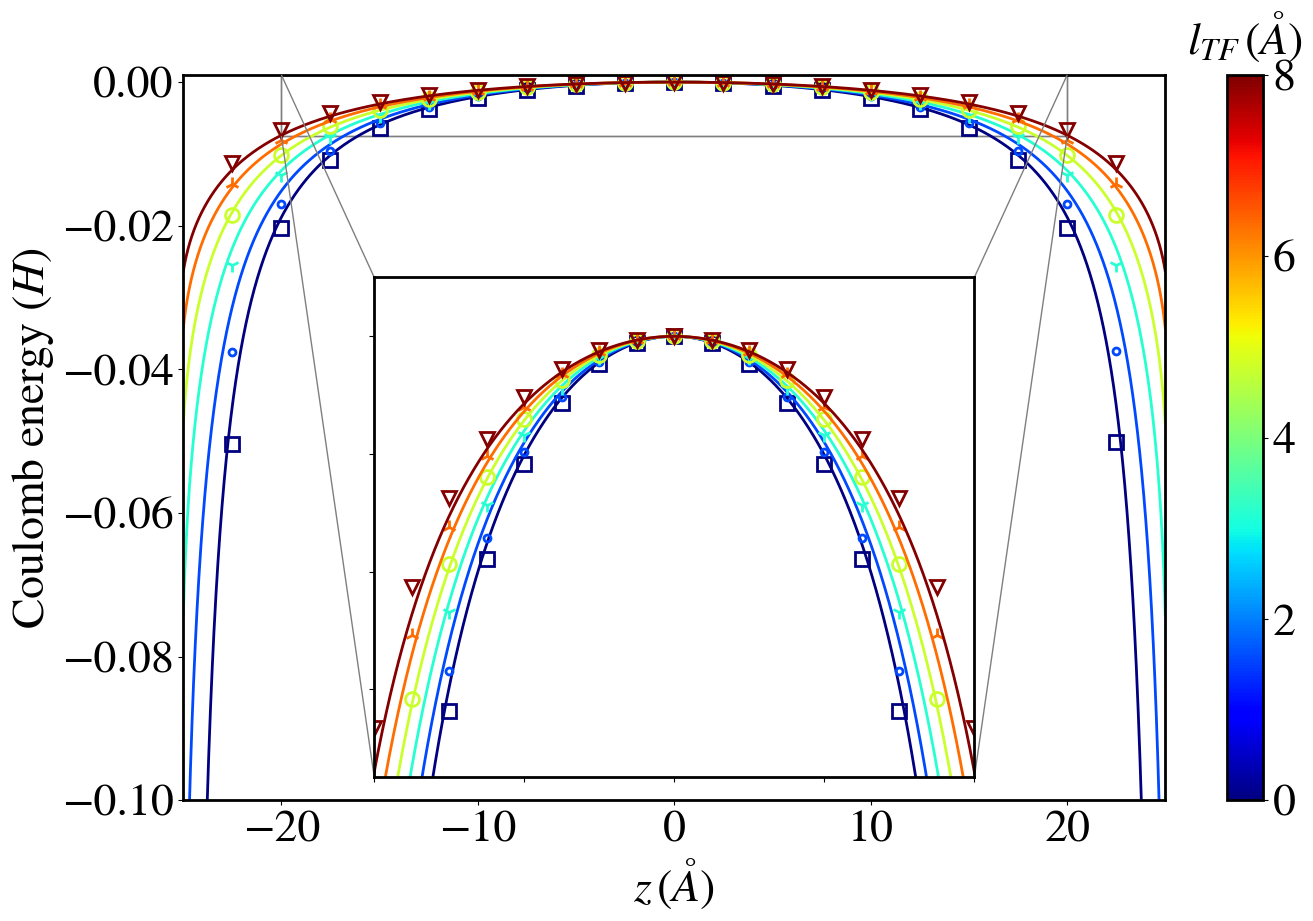}
	\caption{Electrostatic energy of a point charge between two electrodes for a graphitic capacitor with 10 graphene planes per electrode with $L_x = 191.80$~\AA\, and $L_y = 196.88$~\AA. Energies were computed for a range of Thomas-Fermi lengths $l_{TF}$ ranging from 0.0 to 8.0~\AA, represented both by different symbols and by different colors indicated in the colorbar. 
	The simulation results (symbols) are compared to the analytical calculation (straight lines) derived from Kaiser {\it et al.}~\cite{kaiser2017a}. The energies have been shifted to be equal to zero when the ion is at the center of the box.}
\label{fig:capacitance}
\end{figure}

We consider a model system consisting of a unit point charge between two graphite electrodes held at constant potential difference $\Delta \Psi = 0$~V, and compute the energy of the system as a function of the charge position $z$. The energy of a particle of charge $q$ interacting with an infinite continuous Thomas-Fermi metal with a plane surface has been derived in Ref~\citenum{kaiser2017a}, which also provides the following analytical ansatz
\begin{widetext}
\begin{eqnarray}
U_{TF}(z)&=&-\dfrac{q^{2}}{16\pi\epsilon_0z} \left[ 1- \right. \\ \nonumber
& & \left. \dfrac{13.8879(k_{TF}z)^{3} + 37.4625(k_{TF}z)^{2} + 18.6940(k_{TF}z) + 1}{27.8648(k_{TF}z)^{4} + 73.0987(k_{TF}z)^{3} + 70.3460(k_{TF}z)^{2} + 20.6754(k_{TF}z) + 1}\right]
\end{eqnarray}
\end{widetext}
\noindent
where $k_{TF}=1/l_{TF}$ and $z$ is the distance from the surface. Note that for large $k_{TF}$, we recover the result obtained from the image charge method~\cite{reed2007a}. To adapt this expression to our setup, we assumed the lateral dimensions of the box to be large enough to neglect the effect of the periodic images and the two electrodes to be decorrelated in order to simply sum both contributions. We first consider the position of the charge with respect to the atomic position of the electrodes, leading to the following simple expression for the energy of the capacitor composed by a single charged particle and the two electrodes
\begin{equation}
U_{c}=U_{TF}(z)+U_{TF}(L-z)
\end{equation}

\noindent where $L$ is the distance between the electrodes.  The results are shown on Figure \ref{fig:capacitance}. 
The agreement between the simulations and the analytical calculation is good both at long and short ranges and we recover the energy dependence as a function of the Thomas-Fermi length. However, it should be pointed out that such a good comparison can only be achieved using a box with large lateral dimensions (around 200~\AA) compared to the electrodes separation $L=50$~\AA. This is because, for large distances $z$, the effect of the periodic images of the point charge along the lateral directions $x$ and $y$ on the induced charge distribution within the electrodes cannot be neglected, so that the analytical prediction for an isolated ion is not appropriate. 
In addition, our analytical expression assumes a superposition of the contributions due to each electrode treated independently, neglecting in particular the fact that the total charge induced on each electrode is not a full elementary charge but depends on the position of the point charge. Finally, for short distances, there is a threshold below which the continuum approximation breaks down and the molecular structure of the electrode plays a role.


\section{Capacitances of the gold-like electrodes at 1 and 2~V}

\begin{figure}[hbt!]
\centering
\includegraphics[width=1.0\columnwidth]{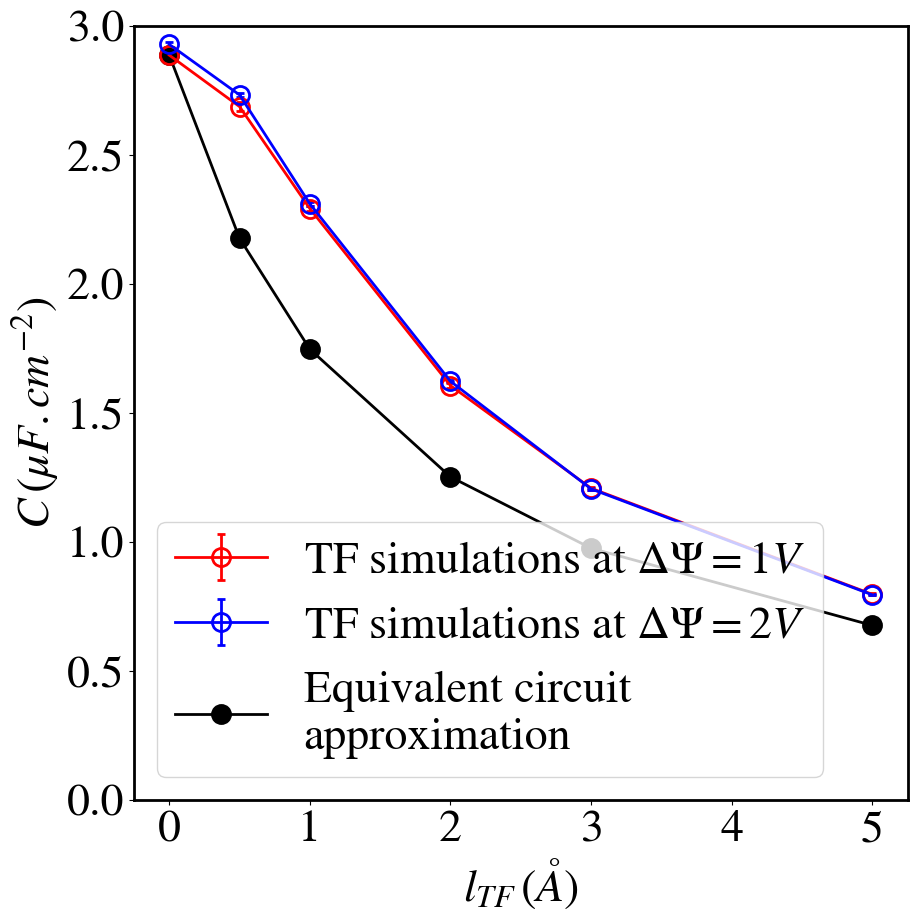}
	\caption{ Integral capacitances computed at applied voltages of 1 and 2~V for the system made of two (100) gold-like electrodes in contact with a NaCl aqueous solution.}
\label{fig:capacitancegold}
\end{figure}

\section{Additional structural characterizations}

\begin{figure}[hbt!]
\centering
\includegraphics[width=1.0\columnwidth]{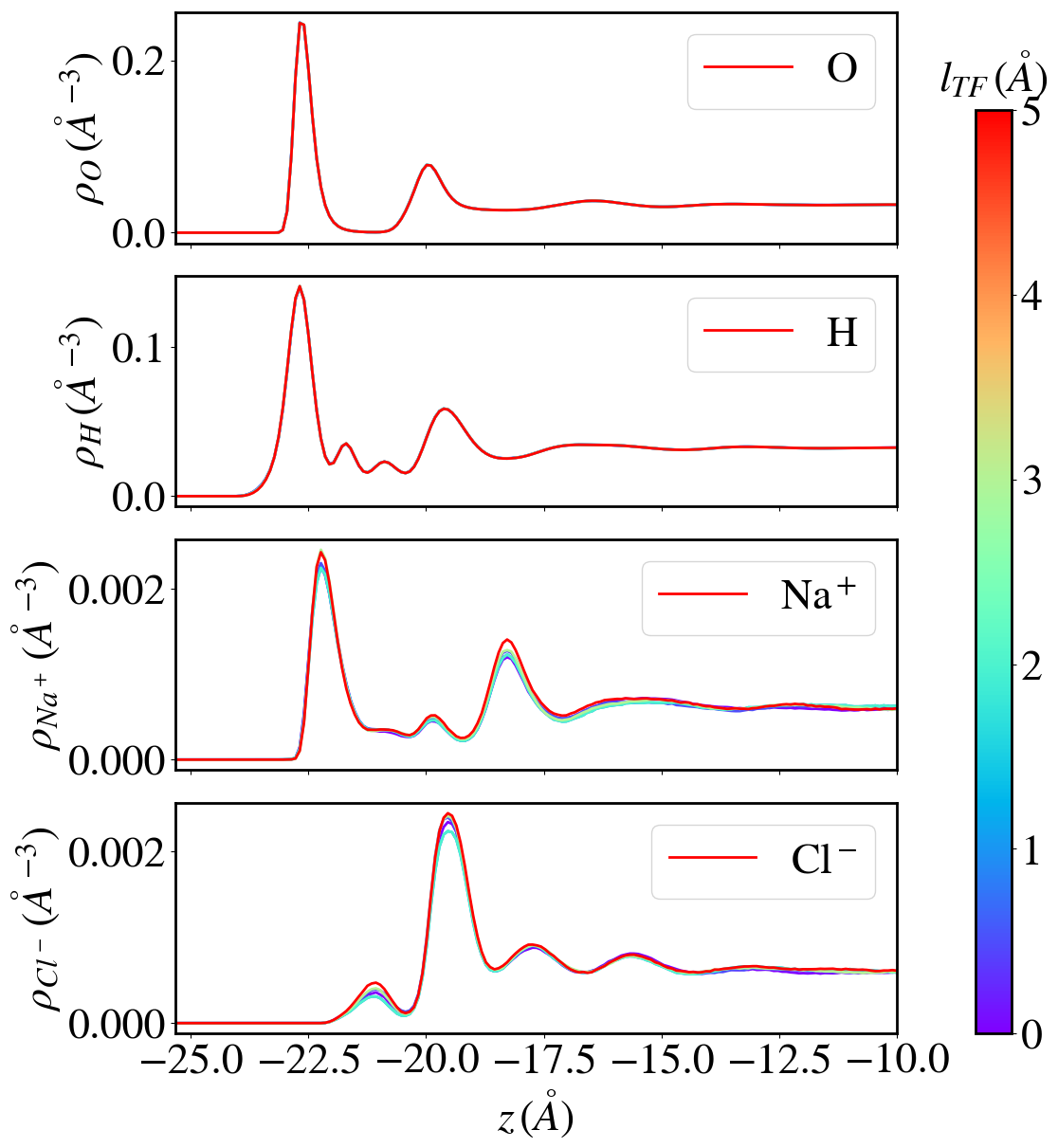}
	\caption{ Atomic density profiles for the O, H, Na$^+$ and Cl$^-$ atoms near the electrode for the systems made of two (100) gold-like electrodes in contact with a NaCl aqueous solution at null potential. The structure is similar for all the $l_{TF}$ values.}
\label{fig:densities0V}
\end{figure}

\begin{figure}[hbt!]
\centering
\includegraphics[width=1.0\columnwidth]{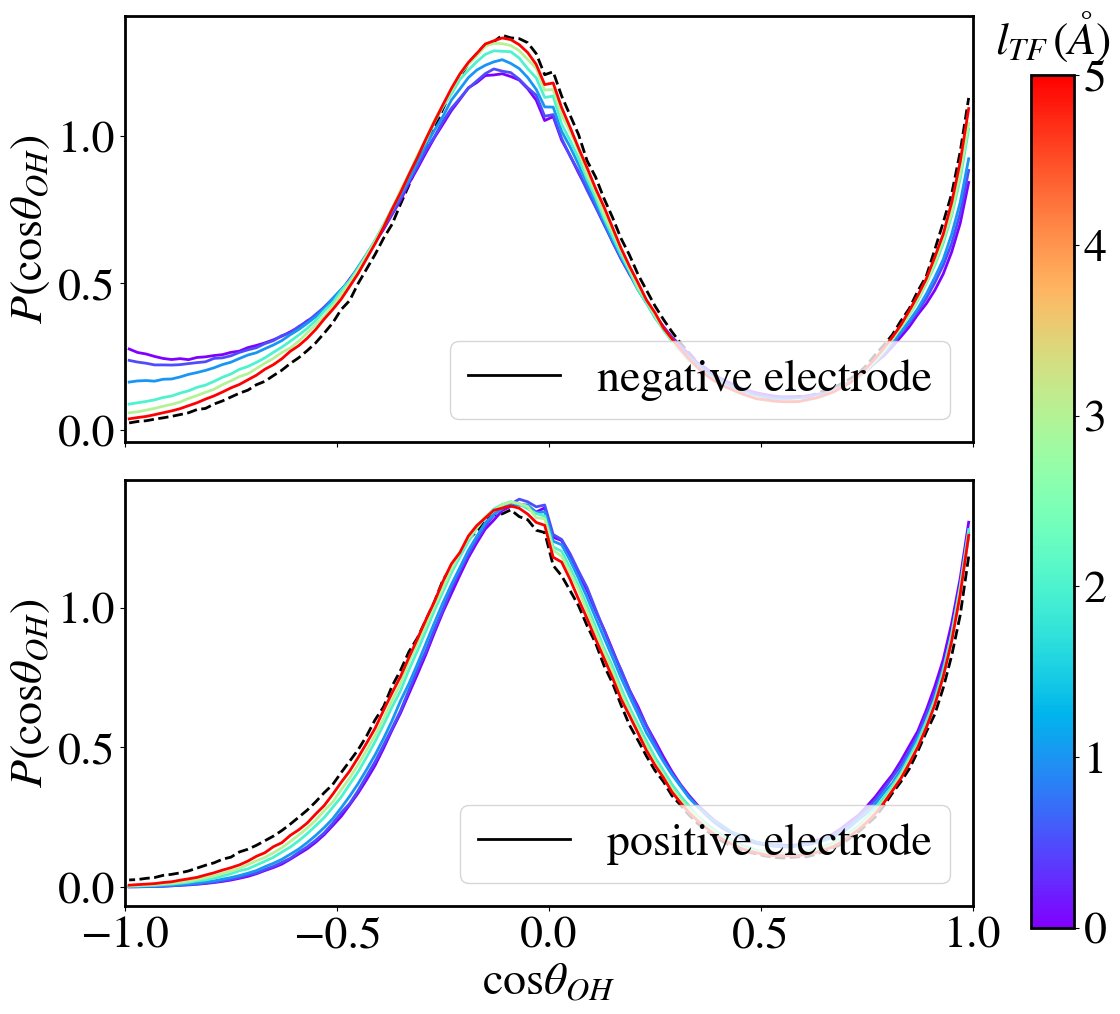}
	\caption{Distribution of the water molecules O-H bonds orientation with respect to the vector normal to the electrode surfaces for an applied potential of 2~V for the whole range of simulated $l_{TF}$ indicated by the colorbar; the distribution for 0~V and $l_{TF}=0.0$~\AA\ is also reported (black dashed lines) as a reference.}
\label{fig:angles}
\end{figure}

\begin{figure}[hbt!]
\centering
\includegraphics[width=1.0\columnwidth]{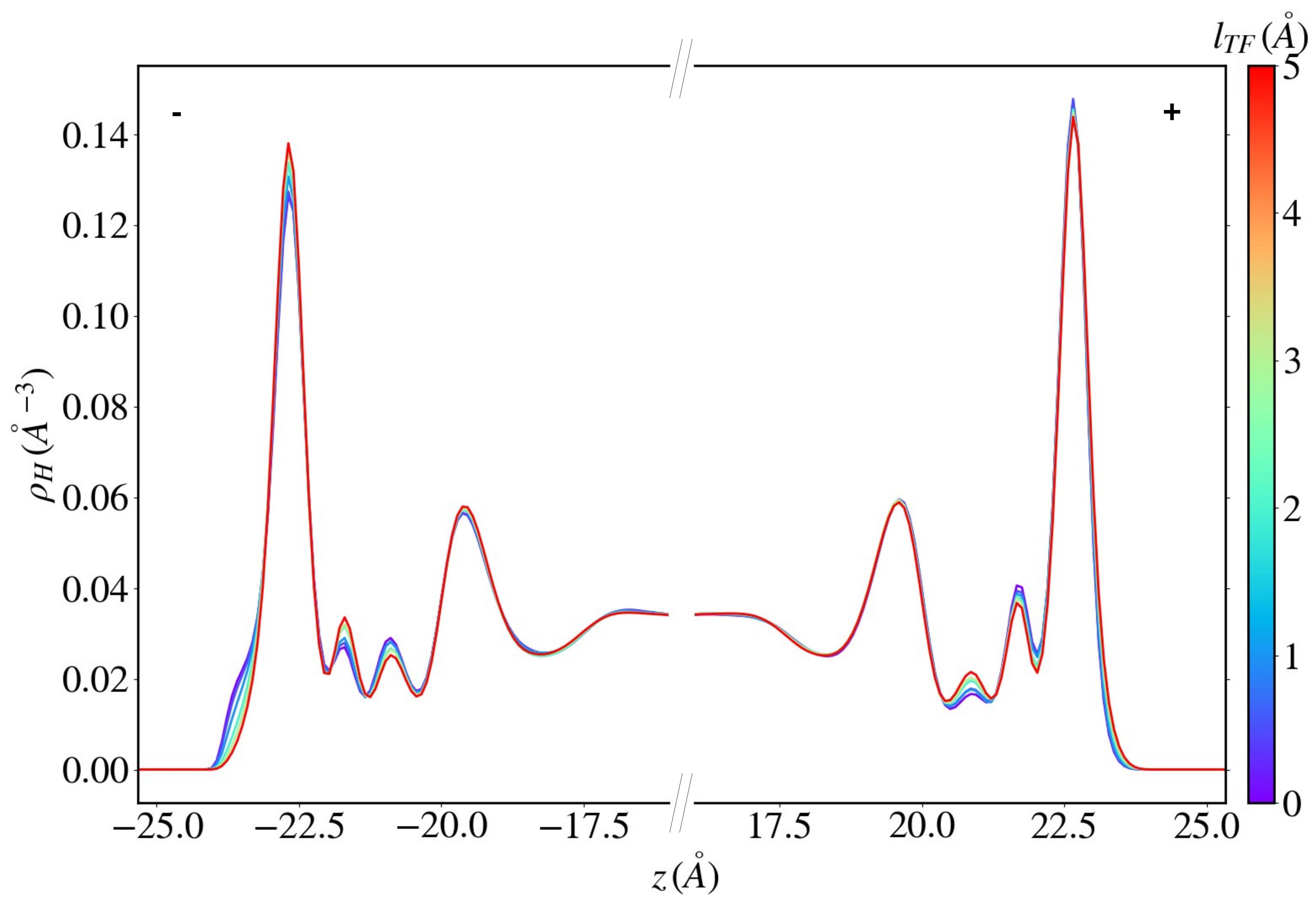}
	\caption{Atomic density profiles for the H atoms for an applied potential of 2~V for the whole range of simulated $l_{TF}$ indicated by the colorbar. The negative (positive) electrode is located at negative (positive) $z$.}
\label{fig:Hdensities}
\end{figure}



\begin{thebibliography}{39}%
\makeatletter
\providecommand \@ifxundefined [1]{%
 \@ifx{#1\undefined}
}%
\providecommand \@ifnum [1]{%
 \ifnum #1\expandafter \@firstoftwo
 \else \expandafter \@secondoftwo
 \fi
}%
\providecommand \@ifx [1]{%
 \ifx #1\expandafter \@firstoftwo
 \else \expandafter \@secondoftwo
 \fi
}%
\providecommand \natexlab [1]{#1}%
\providecommand \enquote  [1]{``#1''}%
\providecommand \bibnamefont  [1]{#1}%
\providecommand \bibfnamefont [1]{#1}%
\providecommand \citenamefont [1]{#1}%
\providecommand \href@noop [0]{\@secondoftwo}%
\providecommand \href [0]{\begingroup \@sanitize@url \@href}%
\providecommand \@href[1]{\@@startlink{#1}\@@href}%
\providecommand \@@href[1]{\endgroup#1\@@endlink}%
\providecommand \@sanitize@url [0]{\catcode `\\12\catcode `\$12\catcode
  `\&12\catcode `\#12\catcode `\^12\catcode `\_12\catcode `\%12\relax}%
\providecommand \@@startlink[1]{}%
\providecommand \@@endlink[0]{}%
\providecommand \url  [0]{\begingroup\@sanitize@url \@url }%
\providecommand \@url [1]{\endgroup\@href {#1}{\urlprefix }}%
\providecommand \urlprefix  [0]{URL }%
\providecommand \Eprint [0]{\href }%
\providecommand \doibase [0]{http://dx.doi.org/}%
\providecommand \selectlanguage [0]{\@gobble}%
\providecommand \bibinfo  [0]{\@secondoftwo}%
\providecommand \bibfield  [0]{\@secondoftwo}%
\providecommand \translation [1]{[#1]}%
\providecommand \BibitemOpen [0]{}%
\providecommand \bibitemStop [0]{}%
\providecommand \bibitemNoStop [0]{.\EOS\space}%
\providecommand \EOS [0]{\spacefactor3000\relax}%
\providecommand \BibitemShut  [1]{\csname bibitem#1\endcsname}%
\let\auto@bib@innerbib\@empty
\bibitem [{\citenamefont {Siepmann}\ and\ \citenamefont
  {Sprik}(1995)}]{siepmann1995a}%
  \BibitemOpen
  \bibfield  {author} {\bibinfo {author} {\bibfnamefont {J.~I.}\ \bibnamefont
  {Siepmann}}\ and\ \bibinfo {author} {\bibfnamefont {M.}~\bibnamefont
  {Sprik}},\ }\bibfield  {title} {\enquote {\bibinfo {title} {{Influence of
  Surface-Topology and Electrostatic Potential on Water Electrode Systems}},}\
  }\href@noop {} {\bibfield  {journal} {\bibinfo  {journal} {J. Chem. Phys.}\
  }\textbf {\bibinfo {volume} {102}},\ \bibinfo {pages} {511--524} (\bibinfo
  {year} {1995})}\BibitemShut {NoStop}%
\bibitem [{\citenamefont {Merlet}\ \emph {et~al.}(2012)\citenamefont {Merlet},
  \citenamefont {Rotenberg}, \citenamefont {Madden}, \citenamefont {Taberna},
  \citenamefont {Simon}, \citenamefont {Gogotsi},\ and\ \citenamefont
  {Salanne}}]{merlet2012a}%
  \BibitemOpen
  \bibfield  {author} {\bibinfo {author} {\bibfnamefont {C.}~\bibnamefont
  {Merlet}}, \bibinfo {author} {\bibfnamefont {B.}~\bibnamefont {Rotenberg}},
  \bibinfo {author} {\bibfnamefont {P.~A.}\ \bibnamefont {Madden}}, \bibinfo
  {author} {\bibfnamefont {P.-L.}\ \bibnamefont {Taberna}}, \bibinfo {author}
  {\bibfnamefont {P.}~\bibnamefont {Simon}}, \bibinfo {author} {\bibfnamefont
  {Y.}~\bibnamefont {Gogotsi}}, \ and\ \bibinfo {author} {\bibfnamefont
  {M.}~\bibnamefont {Salanne}},\ }\bibfield  {title} {\enquote {\bibinfo
  {title} {{On the Molecular Origin of Supercapacitance in Nanoporous Carbon
  Electrodes}},}\ }\href@noop {} {\bibfield  {journal} {\bibinfo  {journal}
  {Nat. Mater.}\ }\textbf {\bibinfo {volume} {11}},\ \bibinfo {pages}
  {306--310} (\bibinfo {year} {2012})}\BibitemShut {NoStop}%
\bibitem [{\citenamefont {Bi}\ \emph {et~al.}(2020)\citenamefont {Bi},
  \citenamefont {Banda}, \citenamefont {Chen}, \citenamefont {Niu},
  \citenamefont {Chen}, \citenamefont {Wu}, \citenamefont {Wang}, \citenamefont
  {Wang}, \citenamefont {Feng}, \citenamefont {Chen}, \citenamefont {Dinca},
  \citenamefont {Kornyshev},\ and\ \citenamefont {Feng}}]{bi2020a}%
  \BibitemOpen
  \bibfield  {author} {\bibinfo {author} {\bibfnamefont {S.}~\bibnamefont
  {Bi}}, \bibinfo {author} {\bibfnamefont {H.}~\bibnamefont {Banda}}, \bibinfo
  {author} {\bibfnamefont {M.}~\bibnamefont {Chen}}, \bibinfo {author}
  {\bibfnamefont {L.}~\bibnamefont {Niu}}, \bibinfo {author} {\bibfnamefont
  {M.}~\bibnamefont {Chen}}, \bibinfo {author} {\bibfnamefont {T.}~\bibnamefont
  {Wu}}, \bibinfo {author} {\bibfnamefont {J.}~\bibnamefont {Wang}}, \bibinfo
  {author} {\bibfnamefont {R.}~\bibnamefont {Wang}}, \bibinfo {author}
  {\bibfnamefont {J.}~\bibnamefont {Feng}}, \bibinfo {author} {\bibfnamefont
  {T.}~\bibnamefont {Chen}}, \bibinfo {author} {\bibfnamefont {M.}~\bibnamefont
  {Dinca}}, \bibinfo {author} {\bibfnamefont {A.~A.}\ \bibnamefont
  {Kornyshev}}, \ and\ \bibinfo {author} {\bibfnamefont {G.}~\bibnamefont
  {Feng}},\ }\bibfield  {title} {\enquote {\bibinfo {title} {Molecular
  understanding of charge storage and charging dynamics in supercapacitors with
  mof electrodes and ionic liquid electrolytes},}\ }\href@noop {} {\bibfield
  {journal} {\bibinfo  {journal} {Nat. Mater.}\ }\textbf {\bibinfo {volume}
  {19}},\ \bibinfo {pages} {552--558} (\bibinfo {year} {2020})}\BibitemShut
  {NoStop}%
\bibitem [{\citenamefont {Limmer}\ \emph {et~al.}(2013)\citenamefont {Limmer},
  \citenamefont {Willard}, \citenamefont {Madden},\ and\ \citenamefont
  {Chandler}}]{limmer2013b}%
  \BibitemOpen
  \bibfield  {author} {\bibinfo {author} {\bibfnamefont {D.~T.}\ \bibnamefont
  {Limmer}}, \bibinfo {author} {\bibfnamefont {A.~P.}\ \bibnamefont {Willard}},
  \bibinfo {author} {\bibfnamefont {P.}~\bibnamefont {Madden}}, \ and\ \bibinfo
  {author} {\bibfnamefont {D.}~\bibnamefont {Chandler}},\ }\bibfield  {title}
  {\enquote {\bibinfo {title} {Hydration of metal surfaces can be dynamically
  heterogeneous and hydrophobic},}\ }\href@noop {} {\bibfield  {journal}
  {\bibinfo  {journal} {Proc. Natl. Acad. Sci. U.S.A.}\ }\textbf {\bibinfo
  {volume} {110}},\ \bibinfo {pages} {4200--4205} (\bibinfo {year}
  {2013})}\BibitemShut {NoStop}%
\bibitem [{\citenamefont {Scalfi}\ \emph {et~al.}(2020)\citenamefont {Scalfi},
  \citenamefont {Limmer}, \citenamefont {Coretti}, \citenamefont {Bonella},
  \citenamefont {Madden}, \citenamefont {Salanne},\ and\ \citenamefont
  {Rotenberg}}]{scalfi2020a}%
  \BibitemOpen
  \bibfield  {author} {\bibinfo {author} {\bibfnamefont {L.}~\bibnamefont
  {Scalfi}}, \bibinfo {author} {\bibfnamefont {D.~T.}\ \bibnamefont {Limmer}},
  \bibinfo {author} {\bibfnamefont {A.}~\bibnamefont {Coretti}}, \bibinfo
  {author} {\bibfnamefont {S.}~\bibnamefont {Bonella}}, \bibinfo {author}
  {\bibfnamefont {P.~A.}\ \bibnamefont {Madden}}, \bibinfo {author}
  {\bibfnamefont {M.}~\bibnamefont {Salanne}}, \ and\ \bibinfo {author}
  {\bibfnamefont {B.}~\bibnamefont {Rotenberg}},\ }\bibfield  {title} {\enquote
  {\bibinfo {title} {Charge fluctuations from molecular simulations in the
  constant-potential ensemble},}\ }\href@noop {} {\bibfield  {journal}
  {\bibinfo  {journal} {Phys. Chem. Chem. Phys.}\ }\textbf {\bibinfo {volume}
  {22}},\ \bibinfo {pages} {10480--10489} (\bibinfo {year} {2020})}\BibitemShut
  {NoStop}%
\bibitem [{\citenamefont {Scalfi}, \citenamefont {Salanne},\ and\ \citenamefont
  {Rotenberg}(2020)}]{scalfi2020arXiv}%
  \BibitemOpen
  \bibfield  {author} {\bibinfo {author} {\bibfnamefont {L.}~\bibnamefont
  {Scalfi}}, \bibinfo {author} {\bibfnamefont {M.}~\bibnamefont {Salanne}}, \
  and\ \bibinfo {author} {\bibfnamefont {B.}~\bibnamefont {Rotenberg}},\
  }\href@noop {} {\enquote {\bibinfo {title} {Molecular simulation of
  electrode-solution interfaces},}\ }\bibinfo {howpublished}
  {\url{https://arxiv.org/abs/2008.11967}} (\bibinfo {year} {2020}),\ \Eprint
  {http://arxiv.org/abs/2008.11967} {arXiv:2008.11967} \BibitemShut {NoStop}%
\bibitem [{\citenamefont {Kornyshev}\ and\ \citenamefont
  {Vorotyntsev}(1978)}]{kornyshev1978a}%
  \BibitemOpen
  \bibfield  {author} {\bibinfo {author} {\bibfnamefont {A.~A.}\ \bibnamefont
  {Kornyshev}}\ and\ \bibinfo {author} {\bibfnamefont {M.~A.}\ \bibnamefont
  {Vorotyntsev}},\ }\bibfield  {title} {\enquote {\bibinfo {title} {Analytic
  expression for the potential energy of atest charge bounded by solid state
  plasma},}\ }\href@noop {} {\bibfield  {journal} {\bibinfo  {journal} {J.
  Phys. C: Solid State Phys.}\ }\textbf {\bibinfo {volume} {11}},\ \bibinfo
  {pages} {L691--L694} (\bibinfo {year} {1978})}\BibitemShut {NoStop}%
\bibitem [{\citenamefont {Kornyshev}, \citenamefont {Schmickler},\ and\
  \citenamefont {Vorotyntsev}(1982)}]{kornyshev1982a}%
  \BibitemOpen
  \bibfield  {author} {\bibinfo {author} {\bibfnamefont {A.~A.}\ \bibnamefont
  {Kornyshev}}, \bibinfo {author} {\bibfnamefont {W.}~\bibnamefont
  {Schmickler}}, \ and\ \bibinfo {author} {\bibfnamefont {M.~A.}\ \bibnamefont
  {Vorotyntsev}},\ }\bibfield  {title} {\enquote {\bibinfo {title} {Nonlocal
  electrostatic approach to the problem of a double layer at a
  metal-electrolyte interface},}\ }\href@noop {} {\bibfield  {journal}
  {\bibinfo  {journal} {Phys. Rev. B}\ }\textbf {\bibinfo {volume} {25}},\
  \bibinfo {pages} {5244--5256} (\bibinfo {year} {1982})}\BibitemShut {NoStop}%
\bibitem [{\citenamefont {Luque}\ and\ \citenamefont
  {Schmickler}(2012)}]{luque2012a}%
  \BibitemOpen
  \bibfield  {author} {\bibinfo {author} {\bibfnamefont {N.~B.}\ \bibnamefont
  {Luque}}\ and\ \bibinfo {author} {\bibfnamefont {W.}~\bibnamefont
  {Schmickler}},\ }\bibfield  {title} {\enquote {\bibinfo {title} {The electric
  double layer on graphite},}\ }\href@noop {} {\bibfield  {journal} {\bibinfo
  {journal} {Electrochim. Acta}\ }\textbf {\bibinfo {volume} {71}},\ \bibinfo
  {pages} {82--85} (\bibinfo {year} {2012})}\BibitemShut {NoStop}%
\bibitem [{\citenamefont {Kornyshev}, \citenamefont {Luque},\ and\
  \citenamefont {Schmickler}(2014)}]{kornyshev2013a}%
  \BibitemOpen
  \bibfield  {author} {\bibinfo {author} {\bibfnamefont {A.~A.}\ \bibnamefont
  {Kornyshev}}, \bibinfo {author} {\bibfnamefont {N.~B.}\ \bibnamefont
  {Luque}}, \ and\ \bibinfo {author} {\bibfnamefont {W.}~\bibnamefont
  {Schmickler}},\ }\bibfield  {title} {\enquote {\bibinfo {title} {Differential
  capacitance of ionic liquid interface with graphite: the story of two double
  layers},}\ }\href@noop {} {\bibfield  {journal} {\bibinfo  {journal} {J.
  Solid State Electrochem.}\ }\textbf {\bibinfo {volume} {18}},\ \bibinfo
  {pages} {1345--1349} (\bibinfo {year} {2014})}\BibitemShut {NoStop}%
\bibitem [{\citenamefont {Comtet}\ \emph {et~al.}(2017)\citenamefont {Comtet},
  \citenamefont {Nigu\`es}, \citenamefont {Kaiser}, \citenamefont {Coasne},
  \citenamefont {Bocquet},\ and\ \citenamefont {Siria}}]{comtet2017a}%
  \BibitemOpen
  \bibfield  {author} {\bibinfo {author} {\bibfnamefont {J.}~\bibnamefont
  {Comtet}}, \bibinfo {author} {\bibfnamefont {A.}~\bibnamefont {Nigu\`es}},
  \bibinfo {author} {\bibfnamefont {V.}~\bibnamefont {Kaiser}}, \bibinfo
  {author} {\bibfnamefont {B.}~\bibnamefont {Coasne}}, \bibinfo {author}
  {\bibfnamefont {L.}~\bibnamefont {Bocquet}}, \ and\ \bibinfo {author}
  {\bibfnamefont {A.}~\bibnamefont {Siria}},\ }\bibfield  {title} {\enquote
  {\bibinfo {title} {Nanoscale capillary freezing of ionic liquids confined
  between metallic interfaces and the role of electronic screening},}\
  }\href@noop {} {\bibfield  {journal} {\bibinfo  {journal} {Nat. Mater.}\
  }\textbf {\bibinfo {volume} {16}},\ \bibinfo {pages} {634--639} (\bibinfo
  {year} {2017})}\BibitemShut {NoStop}%
\bibitem [{\citenamefont {Kaiser}\ \emph {et~al.}(2017)\citenamefont {Kaiser},
  \citenamefont {Comtet}, \citenamefont {Nigu\`es}, \citenamefont {Siria},
  \citenamefont {Coasne},\ and\ \citenamefont {Bocquet}}]{kaiser2017a}%
  \BibitemOpen
  \bibfield  {author} {\bibinfo {author} {\bibfnamefont {V.}~\bibnamefont
  {Kaiser}}, \bibinfo {author} {\bibfnamefont {J.}~\bibnamefont {Comtet}},
  \bibinfo {author} {\bibfnamefont {A.}~\bibnamefont {Nigu\`es}}, \bibinfo
  {author} {\bibfnamefont {A.}~\bibnamefont {Siria}}, \bibinfo {author}
  {\bibfnamefont {B.}~\bibnamefont {Coasne}}, \ and\ \bibinfo {author}
  {\bibfnamefont {L.}~\bibnamefont {Bocquet}},\ }\bibfield  {title} {\enquote
  {\bibinfo {title} {Electrostatic interactions between ions near thomas-fermi
  substrates and the surface energy of ionic crystal at imperfect metals},}\
  }\href@noop {} {\bibfield  {journal} {\bibinfo  {journal} {Faraday Discuss.}\
  }\textbf {\bibinfo {volume} {199}},\ \bibinfo {pages} {129--158} (\bibinfo
  {year} {2017})}\BibitemShut {NoStop}%
\bibitem [{\citenamefont {Thomas}(1927)}]{thomas1927a}%
  \BibitemOpen
  \bibfield  {author} {\bibinfo {author} {\bibfnamefont {L.}~\bibnamefont
  {Thomas}},\ }\bibfield  {title} {\enquote {\bibinfo {title} {The calculation
  of atomic fields},}\ }\href@noop {} {\bibfield  {journal} {\bibinfo
  {journal} {Proc. Cambridge Phil. Roy. Soc.}\ }\textbf {\bibinfo {volume}
  {23}},\ \bibinfo {pages} {542--548} (\bibinfo {year} {1927})}\BibitemShut
  {NoStop}%
\bibitem [{\citenamefont {Fermi}(1927)}]{fermi1927a}%
  \BibitemOpen
  \bibfield  {author} {\bibinfo {author} {\bibfnamefont {E.}~\bibnamefont
  {Fermi}},\ }\bibfield  {title} {\enquote {\bibinfo {title} {Un metodo
  statistico per la determinazione di alcuna priorieta dell'atome},}\
  }\href@noop {} {\bibfield  {journal} {\bibinfo  {journal} {Rend. Accad. Naz.
  Lincei}\ }\textbf {\bibinfo {volume} {6}},\ \bibinfo {pages} {602--607}
  (\bibinfo {year} {1927})}\BibitemShut {NoStop}%
\bibitem [{\citenamefont {Reed}, \citenamefont {Lanning},\ and\ \citenamefont
  {Madden}(2007)}]{reed2007a}%
  \BibitemOpen
  \bibfield  {author} {\bibinfo {author} {\bibfnamefont {S.~K.}\ \bibnamefont
  {Reed}}, \bibinfo {author} {\bibfnamefont {O.~J.}\ \bibnamefont {Lanning}}, \
  and\ \bibinfo {author} {\bibfnamefont {P.~A.}\ \bibnamefont {Madden}},\
  }\bibfield  {title} {\enquote {\bibinfo {title} {{Electrochemical Interface
  Between an Ionic Liquid and a Model Metallic Electrode}},}\ }\href@noop {}
  {\bibfield  {journal} {\bibinfo  {journal} {J. Chem. Phys.}\ }\textbf
  {\bibinfo {volume} {126}},\ \bibinfo {pages} {084704} (\bibinfo {year}
  {2007})}\BibitemShut {NoStop}%
\bibitem [{\citenamefont {Paek}, \citenamefont {Pak},\ and\ \citenamefont
  {Hwang}(2015)}]{paek2015a}%
  \BibitemOpen
  \bibfield  {author} {\bibinfo {author} {\bibfnamefont {E.}~\bibnamefont
  {Paek}}, \bibinfo {author} {\bibfnamefont {A.~J.}\ \bibnamefont {Pak}}, \
  and\ \bibinfo {author} {\bibfnamefont {G.~S.}\ \bibnamefont {Hwang}},\
  }\bibfield  {title} {\enquote {\bibinfo {title} {On the influence of
  polarization effects in predicting the interfacial structure and capacitance
  of graphene-like electrodes in ionic liquids},}\ }\href@noop {} {\bibfield
  {journal} {\bibinfo  {journal} {J. Chem. Phys.}\ }\textbf {\bibinfo {volume}
  {142}},\ \bibinfo {pages} {024701} (\bibinfo {year} {2015})}\BibitemShut
  {NoStop}%
\bibitem [{\citenamefont {Nalewajski}(1984)}]{nalewajski1984a}%
  \BibitemOpen
  \bibfield  {author} {\bibinfo {author} {\bibfnamefont {R.~F.}\ \bibnamefont
  {Nalewajski}},\ }\bibfield  {title} {\enquote {\bibinfo {title}
  {Electrostatic effects in interactions between hard (soft) acids and
  bases},}\ }\href@noop {} {\bibfield  {journal} {\bibinfo  {journal} {J. Am.
  Chem. Soc.}\ }\textbf {\bibinfo {volume} {106}},\ \bibinfo {pages} {944--945}
  (\bibinfo {year} {1984})}\BibitemShut {NoStop}%
\bibitem [{\citenamefont {Mortier}, \citenamefont {Ghosh},\ and\ \citenamefont
  {Shankar}(1986)}]{mortier1986a}%
  \BibitemOpen
  \bibfield  {author} {\bibinfo {author} {\bibfnamefont {W.~J.}\ \bibnamefont
  {Mortier}}, \bibinfo {author} {\bibfnamefont {S.~K.}\ \bibnamefont {Ghosh}},
  \ and\ \bibinfo {author} {\bibfnamefont {S.}~\bibnamefont {Shankar}},\
  }\bibfield  {title} {\enquote {\bibinfo {title}
  {Electronegativity-equalization method for the calculation of atomic charges
  in molecules},}\ }\href@noop {} {\bibfield  {journal} {\bibinfo  {journal}
  {J. Am. Chem. Soc.}\ }\textbf {\bibinfo {volume} {108}},\ \bibinfo {pages}
  {4315--4320} (\bibinfo {year} {1986})}\BibitemShut {NoStop}%
\bibitem [{\citenamefont {Rappe}\ and\ \citenamefont {{Goddard
  III}}(1991)}]{rappe1991a}%
  \BibitemOpen
  \bibfield  {author} {\bibinfo {author} {\bibfnamefont {A.~K.}\ \bibnamefont
  {Rappe}}\ and\ \bibinfo {author} {\bibfnamefont {W.~A.}\ \bibnamefont
  {{Goddard III}}},\ }\bibfield  {title} {\enquote {\bibinfo {title} {Charge
  equilibration for molecular dynamics simulations},}\ }\href@noop {}
  {\bibfield  {journal} {\bibinfo  {journal} {J. Phys. Chem.}\ }\textbf
  {\bibinfo {volume} {95}},\ \bibinfo {pages} {3358--3363} (\bibinfo {year}
  {1991})}\BibitemShut {NoStop}%
\bibitem [{\citenamefont {Onofrio}, \citenamefont {Guzman},\ and\ \citenamefont
  {Strachan}(2015)}]{onofrio2015a}%
  \BibitemOpen
  \bibfield  {author} {\bibinfo {author} {\bibfnamefont {N.}~\bibnamefont
  {Onofrio}}, \bibinfo {author} {\bibfnamefont {D.}~\bibnamefont {Guzman}}, \
  and\ \bibinfo {author} {\bibfnamefont {A.}~\bibnamefont {Strachan}},\
  }\bibfield  {title} {\enquote {\bibinfo {title} {Atomic origin of ultrafast
  resistance switching in nanoscale electrometallization cells},}\ }\href@noop
  {} {\bibfield  {journal} {\bibinfo  {journal} {Nat. Mater.}\ }\textbf
  {\bibinfo {volume} {14}},\ \bibinfo {pages} {440--446} (\bibinfo {year}
  {2015})}\BibitemShut {NoStop}%
\bibitem [{\citenamefont {Buraschi}, \citenamefont {Sansotta},\ and\
  \citenamefont {Zahn}(2020)}]{buraschi2020a}%
  \BibitemOpen
  \bibfield  {author} {\bibinfo {author} {\bibfnamefont {M.}~\bibnamefont
  {Buraschi}}, \bibinfo {author} {\bibfnamefont {S.}~\bibnamefont {Sansotta}},
  \ and\ \bibinfo {author} {\bibfnamefont {D.}~\bibnamefont {Zahn}},\
  }\bibfield  {title} {\enquote {\bibinfo {title} {Polarization effects in
  dynamic interfaces of platinum electrodes and ionic liquid phases: A
  molecular dynamics study},}\ }\href@noop {} {\bibfield  {journal} {\bibinfo
  {journal} {J. Phys. Chem. C}\ }\textbf {\bibinfo {volume} {124}},\ \bibinfo
  {pages} {2002--2007} (\bibinfo {year} {2020})}\BibitemShut {NoStop}%
\bibitem [{\citenamefont {York}\ and\ \citenamefont {Yang}(1996)}]{york1996a}%
  \BibitemOpen
  \bibfield  {author} {\bibinfo {author} {\bibfnamefont {D.~M.}\ \bibnamefont
  {York}}\ and\ \bibinfo {author} {\bibfnamefont {W.}~\bibnamefont {Yang}},\
  }\bibfield  {title} {\enquote {\bibinfo {title} {A chemical potential
  equalization method for molecular simulations},}\ }\href {\doibase
  10.1063/1.470886} {\bibfield  {journal} {\bibinfo  {journal} {The Journal of
  Chemical Physics}\ }\textbf {\bibinfo {volume} {104}},\ \bibinfo {pages}
  {159--172} (\bibinfo {year} {1996})},\ \bibinfo {note} {publisher: American
  Institute of Physics}\BibitemShut {NoStop}%
\bibitem [{\citenamefont {Nakano}\ and\ \citenamefont
  {Sato}(2019)}]{nakano2019a}%
  \BibitemOpen
  \bibfield  {author} {\bibinfo {author} {\bibfnamefont {H.}~\bibnamefont
  {Nakano}}\ and\ \bibinfo {author} {\bibfnamefont {H.}~\bibnamefont {Sato}},\
  }\bibfield  {title} {\enquote {\bibinfo {title} {A chemical potential
  equalization approach to constant potential polarizable electrodes for
  electrochemical-cell simulations},}\ }\href@noop {} {\bibfield  {journal}
  {\bibinfo  {journal} {J. Chem. Phys.}\ }\textbf {\bibinfo {volume} {151}},\
  \bibinfo {pages} {164123} (\bibinfo {year} {2019})}\BibitemShut {NoStop}%
\bibitem [{\citenamefont {Pastewka}\ \emph {et~al.}(2011)\citenamefont
  {Pastewka}, \citenamefont {J\"arvi}, \citenamefont {Mayrhofer},\ and\
  \citenamefont {Moseler}}]{pastewka2011a}%
  \BibitemOpen
  \bibfield  {author} {\bibinfo {author} {\bibfnamefont {L.}~\bibnamefont
  {Pastewka}}, \bibinfo {author} {\bibfnamefont {T.~T.}\ \bibnamefont
  {J\"arvi}}, \bibinfo {author} {\bibfnamefont {L.}~\bibnamefont {Mayrhofer}},
  \ and\ \bibinfo {author} {\bibfnamefont {M.}~\bibnamefont {Moseler}},\
  }\bibfield  {title} {\enquote {\bibinfo {title} {Charge-transfer model for
  carbonaceous electrodes in polar environments},}\ }\href@noop {} {\bibfield
  {journal} {\bibinfo  {journal} {Phys. Rev. B}\ }\textbf {\bibinfo {volume}
  {83}},\ \bibinfo {pages} {165418} (\bibinfo {year} {2011})}\BibitemShut
  {NoStop}%
\bibitem [{\citenamefont {Lang}\ and\ \citenamefont {Kohn}(1973)}]{lang1973a}%
  \BibitemOpen
  \bibfield  {author} {\bibinfo {author} {\bibfnamefont {N.~D.}\ \bibnamefont
  {Lang}}\ and\ \bibinfo {author} {\bibfnamefont {W.}~\bibnamefont {Kohn}},\
  }\bibfield  {title} {\enquote {\bibinfo {title} {Theory of metal surfaces:
  induced surface charge and image potential},}\ }\href@noop {} {\bibfield
  {journal} {\bibinfo  {journal} {Phys. Rev. B}\ }\textbf {\bibinfo {volume}
  {7}},\ \bibinfo {pages} {3541--3550} (\bibinfo {year} {1973})}\BibitemShut
  {NoStop}%
\bibitem [{\citenamefont {Smith}, \citenamefont {Chen},\ and\ \citenamefont
  {Weinert}(1989)}]{smith1989a}%
  \BibitemOpen
  \bibfield  {author} {\bibinfo {author} {\bibfnamefont {N.~V.}\ \bibnamefont
  {Smith}}, \bibinfo {author} {\bibfnamefont {C.~T.}\ \bibnamefont {Chen}}, \
  and\ \bibinfo {author} {\bibfnamefont {M.}~\bibnamefont {Weinert}},\
  }\bibfield  {title} {\enquote {\bibinfo {title} {Distance of the image plane
  from metal surfaces},}\ }\href@noop {} {\bibfield  {journal} {\bibinfo
  {journal} {Phys. Rev. B}\ }\textbf {\bibinfo {volume} {40}},\ \bibinfo
  {pages} {7565--7573} (\bibinfo {year} {1989})}\BibitemShut {NoStop}%
\bibitem [{\citenamefont {Gerischer}(1985)}]{gerischer1985a}%
  \BibitemOpen
  \bibfield  {author} {\bibinfo {author} {\bibfnamefont {H.}~\bibnamefont
  {Gerischer}},\ }\bibfield  {title} {\enquote {\bibinfo {title} {An
  interpretation of the double layer capacity of graphite electrodes in
  relation to the density of states at the {Fermi} level},}\ }\href@noop {}
  {\bibfield  {journal} {\bibinfo  {journal} {J. Phys. Chem.}\ }\textbf
  {\bibinfo {volume} {89}},\ \bibinfo {pages} {4249--4251} (\bibinfo {year}
  {1985})}\BibitemShut {NoStop}%
\bibitem [{\citenamefont {Pak}, \citenamefont {Paek},\ and\ \citenamefont
  {Hwang}(2013)}]{pak2013a}%
  \BibitemOpen
  \bibfield  {author} {\bibinfo {author} {\bibfnamefont {A.~J.}\ \bibnamefont
  {Pak}}, \bibinfo {author} {\bibfnamefont {E.}~\bibnamefont {Paek}}, \ and\
  \bibinfo {author} {\bibfnamefont {G.~S.}\ \bibnamefont {Hwang}},\ }\bibfield
  {title} {\enquote {\bibinfo {title} {Relative contributions of quantum and
  double layer capacitance toward the supercapacitor performance of carbon
  nanotubes in an ionic liquid},}\ }\href@noop {} {\bibfield  {journal}
  {\bibinfo  {journal} {Phys. Chem. Chem. Phys.}\ }\textbf {\bibinfo {volume}
  {15}},\ \bibinfo {pages} {19741--19747} (\bibinfo {year} {2013})}\BibitemShut
  {NoStop}%
\bibitem [{\citenamefont {Salanne}\ \emph {et~al.}(2016)\citenamefont
  {Salanne}, \citenamefont {Rotenberg}, \citenamefont {Naoi}, \citenamefont
  {Kaneko}, \citenamefont {Taberna}, \citenamefont {Grey}, \citenamefont
  {Dunn},\ and\ \citenamefont {Simon}}]{salanne2016a}%
  \BibitemOpen
  \bibfield  {author} {\bibinfo {author} {\bibfnamefont {M.}~\bibnamefont
  {Salanne}}, \bibinfo {author} {\bibfnamefont {B.}~\bibnamefont {Rotenberg}},
  \bibinfo {author} {\bibfnamefont {K.}~\bibnamefont {Naoi}}, \bibinfo {author}
  {\bibfnamefont {K.}~\bibnamefont {Kaneko}}, \bibinfo {author} {\bibfnamefont
  {P.-L.}\ \bibnamefont {Taberna}}, \bibinfo {author} {\bibfnamefont {C.~P.}\
  \bibnamefont {Grey}}, \bibinfo {author} {\bibfnamefont {B.}~\bibnamefont
  {Dunn}}, \ and\ \bibinfo {author} {\bibfnamefont {P.}~\bibnamefont {Simon}},\
  }\bibfield  {title} {\enquote {\bibinfo {title} {{Efficient Storage
  Mechanisms for Building Better Supercapacitors}},}\ }\href@noop {} {\bibfield
   {journal} {\bibinfo  {journal} {Nat. Energy}\ }\textbf {\bibinfo {volume}
  {1}},\ \bibinfo {pages} {16070} (\bibinfo {year} {2016})}\BibitemShut
  {NoStop}%
\bibitem [{\citenamefont {Carrasco}, \citenamefont {Hodgson},\ and\
  \citenamefont {Michaelides}(2012)}]{carrasco2012a}%
  \BibitemOpen
  \bibfield  {author} {\bibinfo {author} {\bibfnamefont {J.}~\bibnamefont
  {Carrasco}}, \bibinfo {author} {\bibfnamefont {A.}~\bibnamefont {Hodgson}}, \
  and\ \bibinfo {author} {\bibfnamefont {A.}~\bibnamefont {Michaelides}},\
  }\bibfield  {title} {\enquote {\bibinfo {title} {A molecular perspective of
  water at metal interfaces},}\ }\href@noop {} {\bibfield  {journal} {\bibinfo
  {journal} {Nat. Mater.}\ }\textbf {\bibinfo {volume} {11}},\ \bibinfo {pages}
  {667--674} (\bibinfo {year} {2012})}\BibitemShut {NoStop}%
\bibitem [{\citenamefont {Fedorov}\ and\ \citenamefont
  {Kornyshev}(2014)}]{fedorov2014a}%
  \BibitemOpen
  \bibfield  {author} {\bibinfo {author} {\bibfnamefont {M.~V.}\ \bibnamefont
  {Fedorov}}\ and\ \bibinfo {author} {\bibfnamefont {A.~A.}\ \bibnamefont
  {Kornyshev}},\ }\bibfield  {title} {\enquote {\bibinfo {title} {Ionic liquids
  at electrified interfaces},}\ }\href@noop {} {\bibfield  {journal} {\bibinfo
  {journal} {Chem. Rev.}\ }\textbf {\bibinfo {volume} {114}},\ \bibinfo {pages}
  {2978---3036} (\bibinfo {year} {2014})}\BibitemShut {NoStop}%
\bibitem [{\citenamefont {Gebbie}\ \emph {et~al.}(2013)\citenamefont {Gebbie},
  \citenamefont {Valtiner}, \citenamefont {Banquy}, \citenamefont {Fox},
  \citenamefont {Henderson},\ and\ \citenamefont
  {Israelachvili}}]{gebbie2013a}%
  \BibitemOpen
  \bibfield  {author} {\bibinfo {author} {\bibfnamefont {M.~A.}\ \bibnamefont
  {Gebbie}}, \bibinfo {author} {\bibfnamefont {M.}~\bibnamefont {Valtiner}},
  \bibinfo {author} {\bibfnamefont {X.}~\bibnamefont {Banquy}}, \bibinfo
  {author} {\bibfnamefont {E.~T.}\ \bibnamefont {Fox}}, \bibinfo {author}
  {\bibfnamefont {W.~A.}\ \bibnamefont {Henderson}}, \ and\ \bibinfo {author}
  {\bibfnamefont {J.~N.}\ \bibnamefont {Israelachvili}},\ }\bibfield  {title}
  {\enquote {\bibinfo {title} {Ionic liquids behave as dilute electrolyte
  solutions},}\ }\href@noop {} {\bibfield  {journal} {\bibinfo  {journal}
  {Proc. Natl. Acad. Sci. U.S.A.}\ }\textbf {\bibinfo {volume} {110}},\
  \bibinfo {pages} {9674--9679} (\bibinfo {year} {2013})}\BibitemShut {NoStop}%
\bibitem [{\citenamefont {Smith}, \citenamefont {Lee},\ and\ \citenamefont
  {Perkin}(2016)}]{smith2016a}%
  \BibitemOpen
  \bibfield  {author} {\bibinfo {author} {\bibfnamefont {A.~M.}\ \bibnamefont
  {Smith}}, \bibinfo {author} {\bibfnamefont {A.~A.}\ \bibnamefont {Lee}}, \
  and\ \bibinfo {author} {\bibfnamefont {S.}~\bibnamefont {Perkin}},\
  }\bibfield  {title} {\enquote {\bibinfo {title} {The electrostatic screening
  length in concentrated electrolytes increases with concentration},}\
  }\href@noop {} {\bibfield  {journal} {\bibinfo  {journal} {J. Phys. Chem.
  Lett.}\ }\textbf {\bibinfo {volume} {7}},\ \bibinfo {pages} {2157--2163}
  (\bibinfo {year} {2016})}\BibitemShut {NoStop}%
\bibitem [{\citenamefont {Nistor}, \citenamefont {Polihronov},\ and\
  \citenamefont {M\"user}(2006)}]{nistor2006a}%
  \BibitemOpen
  \bibfield  {author} {\bibinfo {author} {\bibfnamefont {R.~A.}\ \bibnamefont
  {Nistor}}, \bibinfo {author} {\bibfnamefont {J.~G.}\ \bibnamefont
  {Polihronov}}, \ and\ \bibinfo {author} {\bibfnamefont {M.~H.}\ \bibnamefont
  {M\"user}},\ }\bibfield  {title} {\enquote {\bibinfo {title} {A
  generalization of the charge equilibration method for nonmetallic
  materials},}\ }\href@noop {} {\bibfield  {journal} {\bibinfo  {journal} {J.
  Chem. Phys.}\ }\textbf {\bibinfo {volume} {125}} (\bibinfo {year}
  {2006})}\BibitemShut {NoStop}%
\bibitem [{\citenamefont {Nistor}\ and\ \citenamefont
  {M\"user}(2009)}]{nistor2009a}%
  \BibitemOpen
  \bibfield  {author} {\bibinfo {author} {\bibfnamefont {R.~A.}\ \bibnamefont
  {Nistor}}\ and\ \bibinfo {author} {\bibfnamefont {M.~H.}\ \bibnamefont
  {M\"user}},\ }\bibfield  {title} {\enquote {\bibinfo {title} {Dielectric
  properties of solids in the regular and split-charge equilibration
  formalisms},}\ }\href@noop {} {\bibfield  {journal} {\bibinfo  {journal}
  {Phys. Rev. B}\ }\textbf {\bibinfo {volume} {79}},\ \bibinfo {pages} {104303}
  (\bibinfo {year} {2009})}\BibitemShut {NoStop}%
\bibitem [{\citenamefont {Marin-Lafl\`eche}\ \emph {et~al.}(2020)\citenamefont
  {Marin-Lafl\`eche}, \citenamefont {Haefele}, \citenamefont {Scalfi},
  \citenamefont {Coretti}, \citenamefont {Dufils}, \citenamefont {Jeanmairet},
  \citenamefont {Reed}, \citenamefont {Serva}, \citenamefont {Berthin},
  \citenamefont {Bacon}, \citenamefont {Bonella}, \citenamefont {Rotenberg},
  \citenamefont {Madden},\ and\ \citenamefont {Salanne}}]{marinlafleche2020a}%
  \BibitemOpen
  \bibfield  {author} {\bibinfo {author} {\bibfnamefont {A.}~\bibnamefont
  {Marin-Lafl\`eche}}, \bibinfo {author} {\bibfnamefont {M.}~\bibnamefont
  {Haefele}}, \bibinfo {author} {\bibfnamefont {L.}~\bibnamefont {Scalfi}},
  \bibinfo {author} {\bibfnamefont {A.}~\bibnamefont {Coretti}}, \bibinfo
  {author} {\bibfnamefont {T.}~\bibnamefont {Dufils}}, \bibinfo {author}
  {\bibfnamefont {G.}~\bibnamefont {Jeanmairet}}, \bibinfo {author}
  {\bibfnamefont {S.}~\bibnamefont {Reed}}, \bibinfo {author} {\bibfnamefont
  {A.}~\bibnamefont {Serva}}, \bibinfo {author} {\bibfnamefont
  {R.}~\bibnamefont {Berthin}}, \bibinfo {author} {\bibfnamefont
  {C.}~\bibnamefont {Bacon}}, \bibinfo {author} {\bibfnamefont
  {S.}~\bibnamefont {Bonella}}, \bibinfo {author} {\bibfnamefont
  {B.}~\bibnamefont {Rotenberg}}, \bibinfo {author} {\bibfnamefont {P.~A.}\
  \bibnamefont {Madden}}, \ and\ \bibinfo {author} {\bibfnamefont
  {M.}~\bibnamefont {Salanne}},\ }\bibfield  {title} {\enquote {\bibinfo
  {title} {Metalwalls: A classical molecular dynamics software dedicated to the
  simulation of electrochemical systems},}\ }\href@noop {} {\bibfield
  {journal} {\bibinfo  {journal} {ChemRxiv preprint}\ }\textbf {\bibinfo
  {volume} {under review}} (\bibinfo {year} {2020})}\BibitemShut {NoStop}%
\bibitem [{\citenamefont {Berendsen}, \citenamefont {Grigera},\ and\
  \citenamefont {Straatsma}(1987)}]{berendsen1987a}%
  \BibitemOpen
  \bibfield  {author} {\bibinfo {author} {\bibfnamefont {H.~J.~C.}\
  \bibnamefont {Berendsen}}, \bibinfo {author} {\bibfnamefont {J.~R.}\
  \bibnamefont {Grigera}}, \ and\ \bibinfo {author} {\bibfnamefont {T.~P.}\
  \bibnamefont {Straatsma}},\ }\bibfield  {title} {\enquote {\bibinfo {title}
  {{The Missing Term in Effective Pair Potentials}},}\ }\href@noop {}
  {\bibfield  {journal} {\bibinfo  {journal} {J. Phys. Chem.}\ }\textbf
  {\bibinfo {volume} {91}},\ \bibinfo {pages} {6269--6271} (\bibinfo {year}
  {1987})}\BibitemShut {NoStop}%
\bibitem [{\citenamefont {Dang}(1995)}]{dang1995c}%
  \BibitemOpen
  \bibfield  {author} {\bibinfo {author} {\bibfnamefont {L.~X.}\ \bibnamefont
  {Dang}},\ }\href@noop {} {\bibfield  {journal} {\bibinfo  {journal} {J. Am.
  Chem. Soc.}\ }\textbf {\bibinfo {volume} {117}},\ \bibinfo {pages} {6954}
  (\bibinfo {year} {1995})}\BibitemShut {NoStop}%
\bibitem [{\citenamefont {Berg}, \citenamefont {Peter},\ and\ \citenamefont
  {Johnston}(2017)}]{berg2017a}%
  \BibitemOpen
  \bibfield  {author} {\bibinfo {author} {\bibfnamefont {A.}~\bibnamefont
  {Berg}}, \bibinfo {author} {\bibfnamefont {C.}~\bibnamefont {Peter}}, \ and\
  \bibinfo {author} {\bibfnamefont {K.}~\bibnamefont {Johnston}},\ }\href@noop
  {} {\bibfield  {journal} {\bibinfo  {journal} {J. Chem. Theory Comput.}\
  }\textbf {\bibinfo {volume} {13}},\ \bibinfo {pages} {5610} (\bibinfo {year}
  {2017})}\BibitemShut {NoStop}%
\end{thebibliography}

%
\end{document}